\documentclass[journal]{IEEEtran}
\usepackage{amssymb}

\usepackage{array}
\usepackage{booktabs}
\usepackage{tabularx}

\usepackage{cite}

\usepackage{dirtytalk}
\usepackage{relsize}
\usepackage{exscale}
\usepackage{mathtools}

\ifCLASSINFOpdf
\else
\fi
\usepackage{stackengine}
\usepackage{amsmath}
\usepackage{amsfonts}
\usepackage{siunitx}

\makeatletter
\let\MYcaption\@makecaption
\makeatother

\usepackage{silence}
\usepackage[font=footnotesize]{subcaption}

\makeatletter
\let\@makecaption\MYcaption
\makeatother

\usepackage{multirow}

\usepackage{amsthm}
\usepackage{graphicx}
\graphicspath{ {./images/} }

\usepackage[section]{placeins}
\usepackage{float}

\usepackage{pgfplots}
\pgfplotsset{compat=newest}
\usetikzlibrary{calc}
\usetikzlibrary{plotmarks}
\usetikzlibrary{arrows.meta}
\usetikzlibrary{external}
\usetikzlibrary{decorations.shapes}

\usepgfplotslibrary{patchplots}
\usepackage{grffile}
\makeatletter
\long\def\ifnodedefined#1#2#3{%
    \@ifundefined{pgf@sh@ns@#1}{#3}{#2}%
}
\makeatother

\usepackage{tikzscale}

\usetikzlibrary{positioning}

\usepackage{xcolor}

\definecolor{myorange}{rgb}{0, 0.5, 0}
\definecolor{myturkis}{rgb}{0.8, 0, 1}
\definecolor{myblue}{rgb}{0, 0, 1}
\definecolor{mycolor1}{rgb}{0.00000,0.44700,0.74100}
\definecolor{mycolor2}{rgb}{0.85000,0.32500,0.09800}
\definecolor{mycolor3}{rgb}{0.30100,0.74500,0.93300}
\definecolor{mycolor4}{rgb}{0.63500,0.07800,0.18400}
\definecolor{mycolor5}{rgb}{0.12941,0.12941,0.12941}
\definecolor{mycolor6}{rgb}{0.49400,0.18400,0.55600}

\begin{document}

\DeclareRobustCommand{\markerleg}[1]{%
  \ensuremath{%
    \vcenter{\hbox{%
      \tikz[
        x=1ex, y=1ex,
        baseline
      ]{
        \path[use as bounding box] (-0.7,-0.5) rectangle (0.7,0.5);
        \path[
          #1,
          draw,
          only marks,
          mark size=2.2pt,
          line width=0.3mm
        ] plot coordinates {(0,0)};
      }%
    }}%
  }%
}
\DeclareRobustCommand{\lineleg}[2][]{%
  \tikz[baseline=1.7ex]
        \draw[line width=1,#1,color=#2] (0,0.3) -- (0.5,0.3);%
}

\newcommand{\annotateChangesStart}[0]{}
\newcommand{\annotateChangesEnd}[0]{}

\twocolumn[
\begin{@twocolumnfalse}

\setcounter{page}{0}

\textbf{Author's pre-print} \\

This work has been submitted to the IEEE for possible publication. Copyright may be transferred without notice, after which this version may no longer be accessible.

\end{@twocolumnfalse}
]

\title{A Manifold-Based Framework for Coupling-Aware Surrogate Optimization of Antenna Arrays Using Characteristic Modes}

%
%

\author{Leonardo~M\"orlein,~\IEEEmembership{Student Member,~IEEE,}
        and~Dirk~Manteuffel,~\IEEEmembership{Member,~IEEE}
\thanks{The authors are with the Institute of Microwave and Wireless Systems, Leibniz University Hannover, Appelstr. 9A, 30167 Hannover, Germany (e-mail:
moerlein@imw.uni-hannover.de; manteuffel@imw.uni-hannover.de). This work was partly funded by the Deutsche Forschungsgemeinschaft (DFG, German Research Foundation) – Project number(s) under grant MA 4981/13-1.}
}

%
%

\markboth{}%
{Mörlein \MakeLowercase{\textit{et al.}}: TBD}
%



\maketitle

\begin{abstract}
A surrogate-based synthesis framework for antenna arrays is presented that incorporates mutual coupling while keeping optimization computationally efficient. The method combines a common characteristic-mode basis, a global modal coupling model, and element-wise generalized scattering matrices (GSMs). Array design variables are formulated and optimized on physically meaningful manifolds, in particular the manifold of unitary symmetric matrices for reciprocal and lossless element GSMs. A staged penalty strategy is used to progressively enforce sidelobe and cross-polarization constraints during multi-beam optimization. The framework is demonstrated for an $8\times8$ left-handed circularly polarized patch phased array with scan behavior in one principal plane. Different degree-of-freedom assignment strategies are compared, showing that constrained non-identical element classes can satisfy stringent pattern requirements where equal-element designs fail. For the demonstrated case, the optimization converges within seconds on a single CPU core, and full-wave verification of the realized arrays confirms the predicted trends, with good agreement for the SLL and useful accuracy for the XPR. The results indicate that the proposed formulation is a practical and scalable route for coupling-aware array synthesis and realization.
\end{abstract}


%
\IEEEpeerreviewmaketitle

\raggedbottom

\section{Introduction}

The design of antenna arrays plays a central role in many electromagnetic (EM) applications. Their performance is strongly influenced by mutual coupling between antenna elements, affecting both radiation characteristics and impedance behavior \cite{bird_mutual_2021}. Because accurate modeling of mutual coupling is complex, simplified models that neglect it are often used in early design stages, followed by manual refinement via full-wave EM simulations. While feasible for simple designs, this approach does not scale to complex arrays with many tunable or decoupling elements.

Optimization-based design offers an alternative, but typically requires repeated evaluation of the EM response over the entire simulation domain \cite{nguyen_multi-level_2014,koziel_performance-based_2019,koziel_multi-objective_2013-1,hoorfar_evolutionary_2007,durso_solving_2007}. Using full-wave simulations for this purpose is computationally expensive, motivating surrogate-based optimization techniques \cite{koziel_antenna_2014}. These replace the high-fidelity model with a computationally cheaper approximation. In microwave circuits, such surrogates are often analytical or circuit-based and thus inexpensive to evaluate \cite{salmi_grating-lobe_2024,salmi_synthesis_2024}. For antennas, however, surrogates commonly rely on coarse-discretized EM simulations \cite{koziel_antenna_2014}, which remain costly. The challenge is amplified for antenna arrays due to their high number of degrees of freedom.

For problems with few parameters, methods such as space mapping \cite{bekasiewicz_design_2014,koziel_microstrip_2014} or adaptive response correction \cite{qian_surrogate-assisted_2022} can be applied. For larger systems, alternative strategies are required. One approach partitions the design variables into subsets affecting radiation or impedance, reducing the problem complexity \cite{koziel_simulation_2013}. Another combines array factor models with correction terms derived from a limited number of full-wave simulations \cite{koziel_expedited_2015,koziel_rapid_2015,koziel_rapid_2018}, with hybrid variants also reported \cite{koziel_design_2013,koziel_phase-spacing_2014}. Other methods accelerate full-wave evaluations via macro-basis functions and interpolation \cite{bui_fast_2016,van_fast_2016}, though these remain problem-specific and typically address only element positioning.

\annotateChangesStart
A complementary approach is to optimize the electromagnetic response of the individual elements directly in a reduced modal representation, rather than prescribing a particular geometric parameterization.\annotateChangesEnd{} In our recent works \cite{morlein_array_2025,morlein_deembedding_2025}, we introduced and investigated several building blocks \annotateChangesStart for this approach, \annotateChangesEnd in particular a decomposition of array synthesis into an abstract modal description and its physical realization. These studies employ a modal coupling model in which radiation is expressed via characteristic modes and mutual coupling is captured through a modal coupling matrix, yielding coupled generalized scattering matrices (GSMs). At the array level, this formulation has complexity comparable to circuit models, suggesting its potential as a fast surrogate. Once optimized, individual elements can be realized separately, reducing synthesis effort.

A first application of the underlying modal-coupling and GSM formulation to a circularly polarized array was presented in \cite{morlein_array_2025}, but the optimization procedure there remained tailored to that specific problem. Additionally, \cite{morlein_deembedding_2025} showed that strong geometric modifications can cause deviations between the model and full-wave results. These limitations were further analyzed in \cite{morlein_relation_2025}, revealing that commonly assumed diagonal scattering matrices are not always appropriate, and that element modifications can be interpreted as modal transformations.

While these studies provided useful building blocks and insight, a general surrogate-based optimization framework was still lacking. In this work, we address this gap by formulating a systematic approach based on the modal coupling model. The design variables, represented by GSMs, are treated as elements of the manifold of unitary symmetric matrices. A cost function encodes the optimization objectives, which may be multi-objective. The optimization is performed using the Manopt toolbox \cite{manopt}, yielding optimal GSMs and excitations. A corresponding physical realization is then constructed. This demonstrates a general, physically grounded, and computationally efficient surrogate-based optimization framework for antenna arrays.

In Section~\ref{sec:definitions}, the mathematical definitions are established. Section~\ref{sec:framework} presents the array synthesis framework that forms the core contribution of this work. Section~\ref{sec:example} demonstrates the application of the framework to a practical array design problem. \annotateChangesStart Section~\ref{sec:literature_comparison} provides a quantitative comparison with existing approaches, and Section~\ref{sec:conclusion} concludes the paper. \annotateChangesEnd

\renewcommand{\arraystretch}{1.35}
\newcolumntype{C}{>{\centering\arraybackslash}X}
\begin{table}
    \centering
    \caption{Overview of the structures used throughout the paper}
    \label{tab:overview_structures}
    \begin{tabularx}{\linewidth}{>{\raggedright\arraybackslash}m{3.5cm}CCC}
        \toprule
        & Initial structure & Synthetic Optimized Structure & Realized Structure \\
        \midrule
        Ports present? & No & Yes & Yes \\
        Scattering quantities in common-basis & $\mathbf{S}_0^{(k)}$ & $\widetilde{\mathbf{\Psi}}^{(k)}$ & $\hat{\mathbf{\Psi}}^{(k)}$ \\
        Scattering quantities in eigenbasis & $\mathbf{S}_0^{(k)}$ & $\widetilde{\mathbf{\Psi}}^{(k)}_\mathrm{eig}$ & $\hat{\mathbf{\Psi}}^{(k)}_\mathrm{eig}$ \\
        Terminated scattering quantities in common-basis & n/a & $\widetilde{\mathbf{S}}_\mathrm{0}^{(k)}$ & $\hat{\mathbf{S}}_\mathrm{0}^{(k)}$ \\
        Terminated scattering quantities in eigenbasis & n/a & $\widetilde{\mathbf{S}}_{\mathrm{0},\mathrm{eig}}^{(k)}$ & $\hat{\mathbf{S}}_{\mathrm{0},\mathrm{eig}}^{(k)}$ \\
        Eigenvalues of the terminated structure & n/a & $\widetilde{\lambda}_n^{(k)}$ & $\hat{\lambda}_n^{(k)}$ \\
        Modal transformation matrices from eigenbasis to common basis & $\mathbf{I}$ & $\widetilde{\mathbf{Q}}^{(k)}$ & $\hat{\mathbf{Q}}^{(k)}$ \\
        Eigencurrents & $\mathbf{I}_{\mathrm{CM},n}^{(k)}$ & n/a & $\hat{\mathbf{I}}_{\mathrm{CM},n}^{(k)}$ \\
        Modal far-fields used in the surrogate & $\mathbf{F}_{\mathrm{CM},n}^{(k)}$ & n/a & n.u. \\
        Modal coupling matrix used in the surrogate & $\mathbf{G}$ & n/a & n.u. \\
        Impedance data required & $\mathbf{Z}^{(k,l)}$ for all $k,l$ & - & $\hat{\mathbf{Z}}^{(k,k)}$ for all $k$ \\
        \bottomrule
    \end{tabularx}
    \par\smallskip
    \footnotesize n/a = not applicable, n.u. = quantity exists but is not used explicitly.
\end{table}

\section{Definitions}
\label{sec:definitions}

The main text uses three different structures for the array problem: the initial port-less structure, the synthetic optimized structure and  the realized structure. Their quantities are summarized in Table~\ref{tab:overview_structures} and introduced in the following; the detailed formulas for the realized structure are deferred to Appendix~\ref{sec:app_terminated}.

\subsection{Initial Structure and Common Modal Basis}
\label{sec:def_initial_mom}
\label{sec:def_initial_cm}

The initial structure is assumed to be port-less and is described by the element-wise method-of-moments system
\begin{equation}
    \sum_l \mathbf{Z}^{(k,l)} \mathbf{I}^{(l)} = \mathbf{V}^{(k)}.
\end{equation}
The impedance blocks are given by
\begin{equation}
    \mathbf{Z}^{(k,l)} = \left[ \int_{\Omega_k} \int_{\Omega_l} \mathbf{\psi}_\mu^{(k)}(\mathbf{r}) \mathbf{G}(\mathbf{r},\mathbf{r}') \mathbf{\psi}_m^{(l)}(\mathbf{r}') dS dS' \right]_{\mu m},
\end{equation}
where $\mathbf{G}(\mathbf{r},\mathbf{r}')$ is the dyadic Green's function of the background medium, $\mathbf{\psi}_\mu^{(k)}$ and $\mathbf{\psi}_m^{(l)}$ are the basis functions defined on the $k$\nobreakdash-th and $l$\nobreakdash-th element, respectively, and $\Omega_k$ and $\Omega_l$ are the surfaces of the $k$\nobreakdash-th and $l$\nobreakdash-th element, respectively.

The right hand side $\mathbf{V}^{(k)}$ is given by:
\begin{equation}
    \mathbf{V}^{(k)} = \left[ \int_{\Omega_k} \mathbf{\psi}_\mu^{(k)}(\mathbf{r}) \mathbf{E}_{\mathrm{inc}}(\mathbf{r}) dS \right]_{\mu},
\end{equation}
where $\mathbf{E}_{\mathrm{inc}}(\mathbf{r})$ is the incident electric field at the $k$\nobreakdash-th element.

The surface current on the $k$\nobreakdash-th element is expanded as
\begin{equation}
    \mathbf{J}^{(k)}(\mathbf{r}^\prime) = \sum_m i_m^{(k)} \mathbf{\psi}_m^{(k)}(\mathbf{r}^\prime),
\end{equation}
with coefficient vector
\begin{equation}
    \mathbf{I}^{(k)} = \begin{bmatrix}
        i_1^{(k)} \\
        i_2^{(k)} \\
        \vdots \\
        i_M^{(k)}
    \end{bmatrix}.
\end{equation}

The characteristic modes \cite{garbacz_generalized_1968,harrington_theory_1971,harrington_characteristic_1972} of the $k$\nobreakdash-th element $\mathbf{I}_{\mathrm{CM},n}^{(k)}$ together with their eigenvalues $\lambda_n^{(k)}$ are obtained from
\begin{equation}
    \operatorname{Im} \mathbf{Z}^{(k,k)} \mathbf{I}_{\mathrm{CM},n}^{(k)} = \lambda_n^{(k)} \operatorname{Re} \mathbf{Z}^{(k,k)} \mathbf{I}_{\mathrm{CM},n}^{(k)},
\end{equation}
with normalization
\begin{equation}
    {\mathbf{I}_{\mathrm{CM},n}^{(k)}}^{\mathrm{T}} \operatorname{Re} \mathbf{Z}^{(k,k)} \mathbf{I}_{\mathrm{CM},n}^{(k)} = 1.
\end{equation}
Collecting the modes in
\begin{equation}
    \mathbf{I}_{\mathrm{CM}}^{(k)} = \begin{bmatrix}
        \mathbf{I}_{\mathrm{CM},1}^{(k)} & \mathbf{I}_{\mathrm{CM},2}^{(k)} & \mathbf{I}_{\mathrm{CM},3}^{(k)} & \cdots
    \end{bmatrix},
\end{equation}
the current is written as
\begin{equation}
    \mathbf{I}^{(k)} = \mathbf{I}_{\mathrm{CM}}^{(k)} \mathbf{f}^{(k)}.
\end{equation}

With $\mathbf{a}^{(k)}$ denoting the incident modal coefficients, the response of the $k$\nobreakdash-th element in the initial structure is given by
\begin{equation}
    \label{eq:scattering_matrix}
    \mathbf{f}^{(k)} = \left(\mathbf{S}_0^{(k)} - \mathbf{I} \right) \mathbf{a}^{(k)}.
\end{equation}
where the modal scattering matrix is given by
\begin{equation}
    \mathbf{S}_0^{(k)} = -\operatorname{diag} \frac{1 - j \lambda_n^{(k)}}{1 + j \lambda_n^{(k)}}.
\end{equation}

The incident modal coefficients are composed of external and coupled contributions:
\begin{equation}
    \mathbf{a}^{(k)} = \mathbf{a}_{\mathrm{ext}}^{(k)} + \sum_{\substack{l=1 \\ l \neq k}}^{K} \mathbf{a}^{(k\leftarrow l)}.
\end{equation}
For the synthesis problem considered here, the array is driven only through the element ports, hence
\begin{equation}
    \mathbf{a}_{\mathrm{ext}}^{(k)} = \mathbf{0}.
\end{equation}

Mutual coupling is accounted for through a modal coupling matrix \cite{morlein_array_2025}. In the characteristic mode basis, the coupling from element $l$ to element $k$ is expressed as
\begin{equation}
\label{eq:G_coupling_CM}
    \mathbf{G}^{(k,l)} = \frac{1}{2} {\mathbf{I}^{(k)\mathrm{T}}_{\mathrm{CM}}} \mathbf{Z}^{(k,l)} {\mathbf{I}^{(l)}_{\mathrm{CM}}},
\end{equation}
yielding the coupled modal coefficients
\begin{equation}
    \mathbf{a}^{(k\leftarrow l)} = \mathbf{G}^{(k,l)} \mathbf{f}^{(l)}.
\end{equation}
For the sake of compactness, all coupling matrices are captured in the global modal coupling matrix
\begin{equation}
    \mathbf{G} = \begin{bmatrix}
        \mathbf{0} & \mathbf{G}^{(1,2)} & \cdots & \mathbf{G}^{(1,K)} \\
        \mathbf{G}^{(2,1)} & \mathbf{0} & \ddots & \vdots \\
        \vdots & \ddots & \ddots & \mathbf{G}^{(K-1,K)} \\
        \mathbf{G}^{(K,1)} & \cdots & \mathbf{G}^{(K,K-1)} & \mathbf{0}
    \end{bmatrix}.
\end{equation}

\subsection{Generalized Scattering Model of a Modified Element}
\label{sec:def_ports}

The modified array keeps the element positions of the initial structure, but replaces the port-less elements by elements of different geometry that may contain ports. The common modal basis of the initial structure is retained; only the local scattering relation is generalized. We assume that each element is modelled by $N$ retained characteristic modes and $P$ ports.

For the $k$\nobreakdash-th element, the incident and outgoing port waves over all $S$ excitation states are collected in
\begin{equation}
    \mathbf{v}^{(k)} = \begin{bmatrix} v_{1,1}^{(k)} & \cdots & v_{1,S}^{(k)} \\ \vdots & \ddots & \vdots \\ v_{P,1}^{(k)} & \cdots & v_{P,S}^{(k)} \end{bmatrix},
\end{equation}
and
\begin{equation}
    \mathbf{w}^{(k)} = \begin{bmatrix} w_{1,1}^{(k)} & \cdots & w_{1,S}^{(k)} \\ \vdots & \ddots & \vdots \\ w_{P,1}^{(k)} & \cdots & w_{P,S}^{(k)} \end{bmatrix}.
\end{equation}

The relationship between the incident and outgoing modal weighting coefficients previously given in \eqref{eq:scattering_matrix} is replaced by the following expressions, which include the contribution of the ports:
\begin{equation}
    \mathbf{f}^{(k)} = \left( \widetilde{\mathbf{S}}^{(k)} - \mathbf{I} \right) \mathbf{a}^{(k)} + \widetilde{\mathbf{T}}^{(k)} \mathbf{v}^{(k)}
\end{equation}
and
\begin{equation}
    \mathbf{w}^{(k)} = \widetilde{\mathbf{R}}^{(k)} \mathbf{a}^{(k)} + \widetilde{\mathbf{\Gamma}}^{(k)} \mathbf{v}^{(k)}.
\end{equation}

The matrix $\widetilde{\mathbf{S}}^{(k)}$ is the scattering matrix, $\widetilde{\mathbf{T}}^{(k)}$ is the transmission matrix, $\widetilde{\mathbf{R}}^{(k)}$ is the receive matrix, and $\widetilde{\mathbf{\Gamma}}^{(k)}$ is the port scattering matrix.

Combining both gives the generalized scattering matrix
\begin{equation}
\label{eq:gsm_generic_element_k}
    \Bigg(\underbrace{\begin{bmatrix}
        \widetilde{\mathbf{S}}^{(k)} & \widetilde{\mathbf{T}}^{(k)} \\
        \widetilde{\mathbf{R}}^{(k)} & \widetilde{\mathbf{\Gamma}}^{(k)}
    \end{bmatrix}}_{\widetilde{\mathbf{\Psi}}^{(k)}} - \begin{bmatrix}
        \mathbf{I} & \mathbf{0} \\
        \mathbf{0} & \mathbf{0}
    \end{bmatrix}\Bigg)
    \begin{bmatrix}
        \mathbf{a}^{(k)} \\
        \mathbf{v}^{(k)}
    \end{bmatrix} =
    \begin{bmatrix}
        \mathbf{f}^{(k)} \\
        \mathbf{w}^{(k)}
    \end{bmatrix}.
\end{equation}

For reciprocal, lossless elements, $\widetilde{\mathbf{\Psi}}^{(k)}$ is symmetric and unitary,
\begin{equation}
    \widetilde{\mathbf{\Psi}}^{(k)\mathrm{H}} \widetilde{\mathbf{\Psi}}^{(k)} = \mathbf{I},
\end{equation}
\begin{equation}
    \widetilde{\mathbf{\Psi}}^{(k)\mathrm{T}} = \widetilde{\mathbf{\Psi}}^{(k)},
\end{equation}
which motivates the manifold
\begin{equation}
    M_\mathrm{US} = \left\{\widetilde{\mathbf{\Psi}} \in \mathbb{C}^{(N+P)\times(N+P)} \mid \widetilde{\mathbf{\Psi}}^\mathrm{H} \widetilde{\mathbf{\Psi}} = \mathbf{I},\; \widetilde{\mathbf{\Psi}}^\mathrm{T} = \widetilde{\mathbf{\Psi}}\right\}.
\end{equation}

\section{Array Synthesis Framework}
\label{sec:framework}

\begin{figure*}
    \centering
    \includegraphics{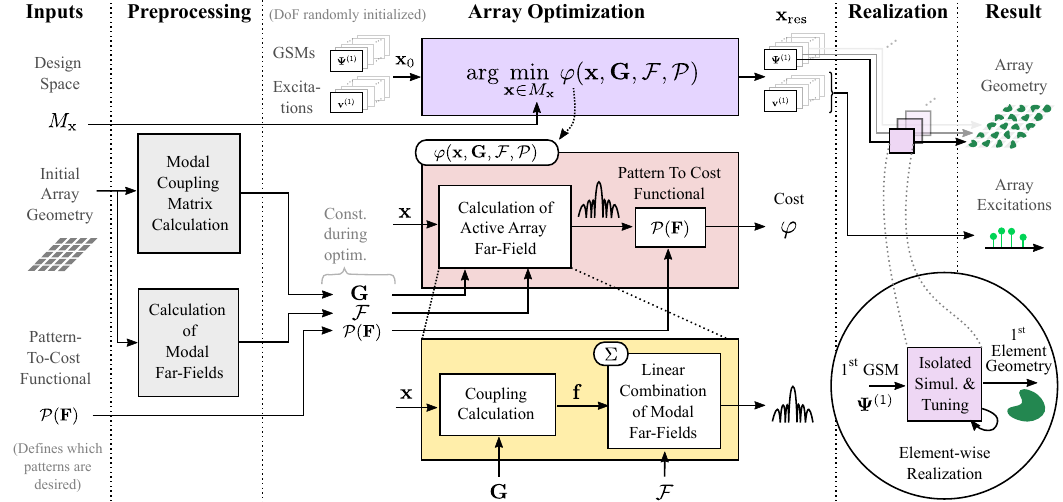}
    \caption{Schematic overview of the proposed approach.}
     \label{fig:overall_approach}
\end{figure*}

The framework separates into four stages: problem specification, preprocessing, array optimization, and physical realization. Fig.~\ref{fig:overall_approach} summarizes this workflow.

\subsection{Problem Specification}

\subsubsection{Initial Geometry}

The initial geometry specifies the element positions and the reference element shapes used to build the common modal basis. During optimization, the element positions remain fixed, while the element geometries are varied only through the surrogate parameters. Consequently, the initial geometry implicitly defines the space of modal currents and patterns that can be excited later on.

\subsubsection{Design Variables and Constraints}

The design variables are the GSMs of the elements and, optionally, the port excitations. They are collected in
\begin{equation}
    \mathbf{x} = \left(\widetilde{\mathbf{\Psi}}^{(1)}, \widetilde{\mathbf{\Psi}}^{(2)}, ..., \widetilde{\mathbf{\Psi}}^{(K)}, \mathbf{v}^{(1)}, \mathbf{v}^{(2)}, ..., \mathbf{v}^{(K)}\right).
\end{equation}

For each element,
\begin{equation}
    \widetilde{\mathbf{\Psi}}^{(k)} \in M_{\mathrm{US}}.
\end{equation}

If several elements are constrained to share one geometry, a degree-of-freedom index $d$ is introduced and one GSM is assigned to each geometry class:
\begin{equation}
    \widetilde{\mathbf{\Psi}}_d \in M_{\mathrm{US}}, \quad d = 1, 2, ..., D.
\end{equation}

This separates the physically meaningful constraints on the element response from the bookkeeping of which array positions share the same design variable. If desired, excitation constraints can be imposed analogously by choosing a suitable manifold for the port excitations.

The resulting feasible set of all optimization variables is denoted by $M_{\mathbf{x}}$, i.e., the product manifold formed by the chosen GSM and excitation constraints.

\subsubsection{Pattern-to-Cost Functional}

The optimization objective is defined on the far-field $\mathbf{F}(\theta, \phi)$ which already includes mutual coupling. For compact notation, let
\begin{equation}
    \xi = (\theta, \phi).
\end{equation}

A beam specification $\sigma$ collects the target direction $\xi_{\mathrm{t}}(\sigma)$, desired and undesired polarization vectors $\mathbf{u}_{\mathrm{D}}(\sigma)$ and $\mathbf{u}_{\mathrm{X}}(\sigma)$, target sidelobe level $\mathrm{SLL}(\sigma)$, target cross-polarization ratio $\mathrm{XPR}(\sigma)$, and the angular sample sets $M_{\mathrm{S}}(\sigma)$ and $M_{\mathrm{X}}(\sigma)$ on which these targets are enforced.

For a single beam, the auxiliary cost functional acting on a single far-field $\mathbf{F}(\xi)$ is
\begin{equation}
\begin{aligned}
    p_{\alpha}(\mathbf{F}, \sigma) = & -|\mathbf{u}_{\mathrm{D}}(\sigma)\mathbf{F}(\xi_{\mathrm{t}}(\sigma))|^2 \\
        & + \alpha \sum_{\xi \in M_{\mathrm{S}}(\sigma)} \gamma\left( \frac{|\mathbf{u}_{\mathrm{D}}(\sigma)\mathbf{F}(\xi)|}{|\mathbf{u}_{\mathrm{D}}(\sigma)\mathbf{F}(\xi_{\mathrm{t}}(\sigma))| \mathrm{SLL}(\sigma)} \right)^2 \\
        & + \alpha \sum_{\xi  \in M_{\mathrm{X}}(\sigma) }\gamma\left( \frac{|\mathbf{u}_{\mathrm{X}}(\sigma)\mathbf{F}(\xi)|}{|\mathbf{u}_{\mathrm{D}}(\sigma)\mathbf{F}(\xi_{\mathrm{t}}(\sigma))| \mathrm{XPR}(\sigma)} \right)^2,
\end{aligned}
\end{equation}
with $ \gamma(x) $ being the penalty function which applies a linear penalty if its argument exceeds 1 and no penalty otherwise:
\begin{equation}
    \gamma(x) = \begin{cases}
        0, & x \leq 1\\
        x - 1, & x > 1
    \end{cases}.
\end{equation}
The parameter $\alpha$ balances main-beam gain against sidelobe and cross-polarization penalties. For $S$ desired beams denoted by $\sigma_s$, the overall cost is
\begin{equation}
\label{eq:pattern_to_cost_functional}
\begin{aligned}
    \mathcal{P}_{\alpha}(\mathbf{F}_1, ... , \mathbf{F}_S) & =  \sum_{s = 1}^S p_{\alpha}(\mathbf{F}_s, \sigma_s), \\
\end{aligned}
\end{equation}
where $\mathbf{F}_s(\xi)$ denotes the far-field of the $s$\nobreakdash-th excitation state.

\subsection{Preprocessing - Model Construction}

Before the optimization starts, the surrogate model is assembled once from the initial structure. This preprocessing stage provides the impedance blocks $\mathbf{Z}^{(k,l)}$, the characteristic-mode bases $\mathbf{I}^{(k)}_{\mathrm{CM}}$, the modal far-fields $\mathbf{F}_{\mathrm{CM},n}^{(k)}(\theta, \phi)$, and the global coupling matrix $\mathbf{G}$. These quantities remain fixed throughout the optimization, while the design variables $\mathbf{x}$ are updated.

\annotateChangesStart
In particular, $\mathbf{G}$ is obtained by projecting the assembled MoM impedance blocks onto the modal bases according to \eqref{eq:G_coupling_CM}; no driven full-array MoM solve is required for this construction. Preprocessing still entails characteristic-mode eigenproblems and modal far-field evaluation. During array optimization, only the reduced coupled GSM system is solved, without additional driven full-wave simulations or a training dataset of modified array configurations. This distinction concerns model construction and array optimization, not the subsequent physical realization and full-wave validation.
\annotateChangesEnd

\subsection{Array Optimization}

For each candidate design, the coupled-element response is computed in the common modal basis. Given the element GSMs, the port excitations, and the precomputed coupling matrix, the outgoing modal coefficients are obtained from the coupled GSM system as outlined in \cite{morlein_array_2025}:
\begin{equation}
    \mathbf{f} = f(\mathbf{x}, \mathbf{G}),
\end{equation}
where $\mathbf{f}$ collects the outgoing modal coefficients of all elements, which already contain all coupling effects:
\begin{equation}
    \mathbf{f} = \begin{bmatrix} \mathbf{f}^{(1)} \\ \mathbf{f}^{(2)} \\ \vdots \\ \mathbf{f}^{(K)} \end{bmatrix} \in \mathbb{C}^{(KN) \times S}.
\end{equation}

Collecting the far-fields of all retained modes in $\mathcal{F}$, define
\begin{equation}
    \mathcal{F} = \begin{bmatrix}
        \mathbf{F}_{\mathrm{CM},n}^{(k)}(\theta, \phi)
    \end{bmatrix}_n^{(k)},
\end{equation}
and the mapping from modal coefficients to far-fields as
\begin{equation}
\begin{aligned}
    \operatorname{FF}(\mathbf{f}, \mathcal{F}) &= \left(\mathbf{F}_1(\theta, \phi), ..., \mathbf{F}_S(\theta, \phi)\right) \\
    &= \left(\sum_{k = 1}^{K} \sum_{n = 1}^{N} \mathbf{F}_{\mathrm{CM},n}^{(k)}(\theta, \phi) f_{n,s}^{(k)}\right)_s.
\end{aligned}
\end{equation}

The optimization problem is given as
\begin{equation}
    \mathbf{x}_{\mathrm{res}} = \arg\min_{\mathbf{x} \in M_{\mathbf{x}}} \varphi(\mathbf{x}, \mathbf{G}, \mathcal{F}, \mathcal{P}).
\end{equation}

With the building blocks above,
\begin{equation}
    \varphi(\mathbf{x}, \mathbf{G}, \mathcal{F}, \mathcal{P}) = \mathcal{P}(\operatorname{FF}(f(\mathbf{x}, \mathbf{G}), \mathcal{F})).
\end{equation}
This decomposition is exactly the reason for separating preprocessing quantities from the optimization variables: the expensive electromagnetic information is frozen in $\mathbf{G}$ and $\mathcal{F}$, while the optimizer only updates $\mathbf{x}$.

\begin{figure}
    \centering
    \includegraphics{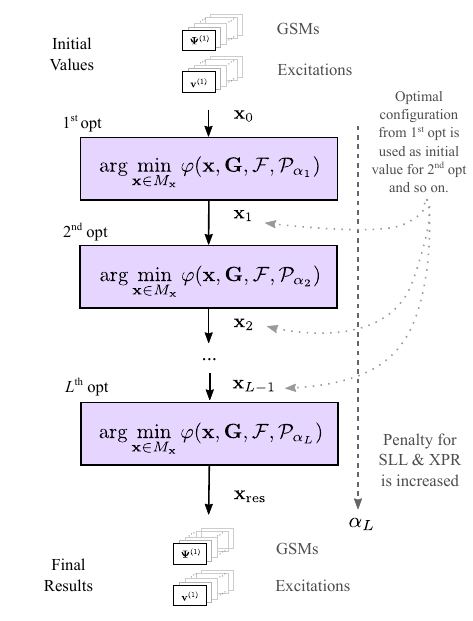}
    \caption{Schematic overview of the proposed approach with repeated optimization with increasing penalty for the SLL and XPR.}
     \label{fig:overall_approach_repeated_opt}
\end{figure}

In practice, better solutions were obtained by solving a sequence of optimization problems with gradually increasing SLL and XPR penalty $\alpha$. The procedure starts with $\alpha = 0$, which only optimizes the gain at the target angle, and then gradually tightens the sidelobe and cross-polarization requirements; the solution of one stage is used to initialize the next one, as illustrated in Fig.~\ref{fig:overall_approach_repeated_opt}. Since the variables are constrained to a manifold, this stage can be implemented with a manifold optimizer such as Manopt.

\subsection{Realization of Optimized Elements}
\label{sec:realization}

After optimization, each element has to be mapped to a realizable geometry and port configuration whose GSM approximates the optimized target. For that purpose, each GSM is partitioned into
\begin{equation}
    \widetilde{\mathbf{\Psi}}^{(k)} = \begin{bmatrix}
    \widetilde{\mathbf{S}}^{(k)} & \widetilde{\mathbf{T}}^{(k)} \\
    \widetilde{\mathbf{R}}^{(k)} & \widetilde{\mathbf{\Gamma}}^{(k)}
\end{bmatrix},
\end{equation}
with the interpretation of scattering, transmit, receive, and port-reflection blocks. To use characteristic modes as a design tool, it is helpful to transform the optimized GSM into a form that has no ports. The antenna is therefore terminated with a load reflection coefficient $\mathbf{\Gamma}_\mathrm{L}$ (usually open-circuit or short-circuit). Its modal scattering matrix is
\begin{equation}
    \widetilde{\mathbf{S}}_0^{(k)} = \widetilde{\mathbf{S}}^{(k)} + \widetilde{\mathbf{T}}^{(k)} (\mathbf{\Gamma}_\mathrm{L}  - \widetilde{\mathbf{\Gamma}}^{(k)})^{-1} \widetilde{\mathbf{R}}^{(k)}.
\end{equation}

As shown in~\cite{morlein_relation_2025}, this matrix is generally not diagonal, so the common modal basis is no longer an eigenbasis of the realized structure. By solving the eigenvalue problem of the terminated antenna
\begin{equation}
    \widetilde{\mathbf{S}}^{(k)}_0 \widetilde{\mathbf{Q}}^{(k)} = \widetilde{\mathbf{Q}}^{(k)} \widetilde{\mathbf{S}}^{(k)}_{0,\mathrm{eig}},
\end{equation}
the eigenvalues $\widetilde{\lambda}_n^{(k)}$ in the diagonal of
\begin{equation}
    \widetilde{\mathbf{S}}^{(k)}_{0,\mathrm{eig}} = \mathrm{diag} \frac{1 - \mathrm{j}\widetilde{\lambda}^{(k)}_n}{1 + \mathrm{j}\widetilde{\lambda}^{(k)}_n}
\end{equation}
and the quasi-orthogonal modal transformation matrix $\widetilde{\mathbf{Q}}^{(k)}$, which maps coefficients from the eigenbasis to the common basis, are obtained. The transmit matrix transformed to the eigenbasis is given by
\begin{equation}
    \widetilde{\mathbf{T}}_{\mathrm{eig}}^{(k)} = \widetilde{\mathbf{Q}}^{(k)\mathrm{T}} \widetilde{\mathbf{T}}^{(k)}.
\end{equation}

The resulting eigenvalues indicate which modal behavior the physical element must realize, the transformation matrix defines the orientation of the modes of the desired structure, and the modal transmit matrix specifies which modes should be excited by the ports. With the help of Appendix~\ref{sec:app_terminated}, a structure can then be designed to realize this modal behavior and excitation. This process is repeated for each element, and the resulting geometries are assembled into the final array.

\subsection{Reference-Plane Phase Shifts for Realization}
\label{sec:invariance_phaseshift}

\begin{figure}[!tb]
\centering
\includegraphics{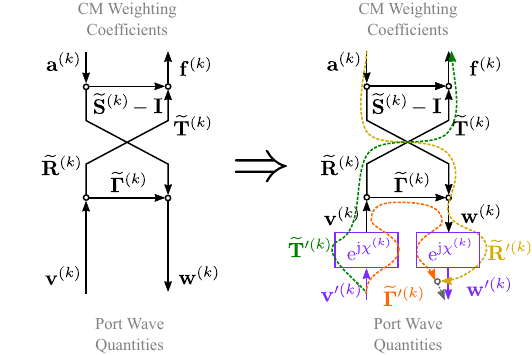}
\caption{Signal flow representation of the $k$-th antenna element with original (left) and phase-shifted reference plane (right). The identical incident ($\mathbf{a}^{(k)}$) and radiated ($\mathbf{f}^{(k)}$) mode coefficients are shared between both configurations.}
\label{fig:element_phaseshift}
\end{figure}

Some optimized eigenvalues $\widetilde{\lambda}_n^{(k)}$ may be difficult to realize physically. A useful degree of freedom is that the cost function is invariant under phase shifts of the element reference planes, so the same far-field can be represented by different GSMs. The corresponding signal-flow interpretation is illustrated in Fig.~\ref{fig:element_phaseshift}.

In terms of the building blocks of the cost function, this invariance reads
\begin{equation}
    f(\mathbf{x}, \mathbf{G}) = f(\mathbf{x}^\prime, \mathbf{G}),
\end{equation}
where $\widetilde{\mathbf{\Psi}}^{\prime(k)}$ and $\mathbf{v}^{\prime(k)}$ inside $\mathbf{x}^\prime$ are obtained by shifting the port reference plane of the $k$\nobreakdash-th element by a phase $\chi^{(k)}$. The transformed quantities are
\begin{equation}
    \widetilde{\mathbf{\Psi}}^{\prime(k)} = \begin{bmatrix}
        \widetilde{\mathbf{S}}^{(k)} & \widetilde{\mathbf{T}}^{(k)} \mathrm{e}^{\mathrm{j}\chi^{(k)}} \\
        \widetilde{\mathbf{R}}^{(k)} \mathrm{e}^{\mathrm{j}\chi^{(k)}} & \widetilde{\mathbf{\Gamma}}^{(k)} \mathrm{e}^{\mathrm{j}2\chi^{(k)}}
    \end{bmatrix}\text{and}\;
    \mathbf{v}^{\prime(k)} = \mathbf{v}^{(k)} \mathrm{e}^{-\mathrm{j}\chi^{(k)}},
\end{equation}
with free parameter $\chi^{(k)} \in [0,2\pi)$. Although this transformation leaves the array response unchanged, it modifies the scattering matrix of the terminated antenna to
\begin{equation}
    \widetilde{\mathbf{S}}^{\prime(k)}_0(\chi^{(k)}) = \widetilde{\mathbf{S}}^{(k)} +  \widetilde{\mathbf{T}}^{(k)} \left(\mathbf{\Gamma}_\mathrm{L} \mathrm{e}^{-\mathrm{j}2\chi^{(k)}}  - \widetilde{\mathbf{\Gamma}}^{(k)} \right)^{-1} \widetilde{\mathbf{R}}^{(k)}.
\end{equation}

Its eigenvalue problem is
\begin{equation}
    \widetilde{\mathbf{S}}^{\prime(k)}_0(\chi^{(k)}) \widetilde{\mathbf{Q}}^{\prime(k)}(\chi^{(k)}) = \widetilde{\mathbf{Q}}^{\prime(k)}(\chi^{(k)}) \widetilde{\mathbf{S}}^{\prime(k)}_{0,\mathrm{eig}}(\chi^{(k)}).
\end{equation}

The diagonal matrix contains the realization-relevant eigenvalues,
\begin{equation}
    \widetilde{\mathbf{S}}^{\prime(k)}_{0,\mathrm{eig}}(\chi^{(k)}) = \mathrm{diag} \frac{1 - \mathrm{j}\widetilde{\lambda}^{\prime(k)}_n(\chi^{(k)})}{1 + \mathrm{j}\widetilde{\lambda}^{\prime(k)}_n(\chi^{(k)})}.
\end{equation}

and the transmit matrix in that eigenbasis is
\begin{equation}
        \widetilde{\mathbf{T}}_{\mathrm{eig}}^{\prime(k)}(\chi^{(k)}) = \widetilde{\mathbf{Q}}^{\prime(k)}(\chi^{(k)})^\mathrm{T} \widetilde{\mathbf{T}}^{(k)} \mathrm{e}^{\mathrm{j}\chi^{(k)}}.
\end{equation}

This freedom can be exploited to shift the optimized design toward a more realizable set of eigenvalues without changing the synthesized far-field.

\begin{figure}
    \centering
    \includegraphics{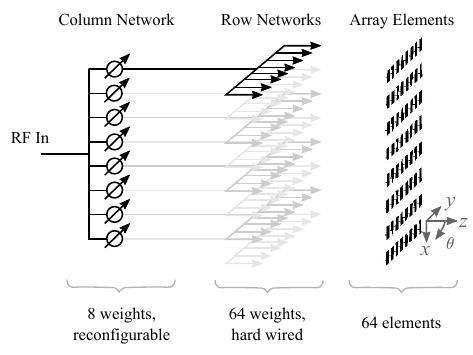}
    \caption{Virtual feed network that should be used to excite the array.}
    \label{fig:example_distribution_network}
\end{figure}

\section{Example}
\label{sec:example}

\begin{figure}
    \centering
    \subfloat[Element Geometry]{\includegraphics{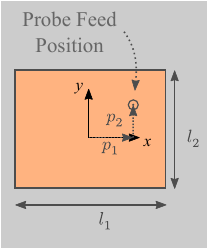}}\hspace{5mm}
    \subfloat[Isolated Element Pattern]{\input{images/example_array_pattern_element_init_iso.tex}}
    \caption{Initial element. The element has been tuned to achieve LHCP in the isolated case, which led to edge lengths of the patch $l_1 = 0.24\lambda_0$ and $l_2 = 0.22\lambda_0$ and the port position $(p_1^{(k)}, p_2^{(k)}) = (0.047\lambda_0, 0.031\lambda_0)$ are shown. The desired LHCP component (\lineleg{mycolor1}) and the undesired RHCP component (\lineleg{mycolor2}) are shown respectively. An arrow marks the XPR at the target angle $\theta_t = 0^\circ$.}
    \label{fig:initial_element_geometry}
\end{figure}

The purpose of the following example is to illustrate the proposed optimization method by applying it to a circularly polarized array design problem that is representative of a practical phased array scenario. The example is intended to demonstrate the proposed method under realistic design requirements, without aiming at a finalized hardware design for a specific application.

\subsection{Problem Statement}

An array of $8\times8$ patch antenna elements on a dielectric substrate with $\varepsilon_\mathrm{r} = 3.66$ is considered. It is designed to achieve left-handed circular polarization (LHCP). The array should be designed as a phased array, which means that it should be able to dynamically scan the beam in different directions. For simplicity, the array is only designed to scan in a single plane at $\phi = 0^\circ$. The array should be able to scan up to $\theta = \pm 60^\circ$ in this plane. A cross polarization ratio $\mathrm{XPR} = \SI{30}{\dB}$ and sidelobe level $\mathrm{SLL} = \SI{-15}{\dB}$ are required. The inter-element spacing is fixed to $D_x = D_y = \lambda_0/2$, and the array is designed on a substrate with a height of $h = \lambda_0/21$.

Since the array is a phased array, its excitation is only partially fixed and can be adjusted for beam steering. Because scanning is required only in one plane, a simple feed network is sufficient (Fig.~\ref{fig:example_distribution_network}). The column network is dynamically reconfigurable for steering, whereas the row network is fixed during operation; its weights are optimized once during design.

\annotateChangesStart
This one-dimensional scan scenario is chosen deliberately because, despite the reduced scan space, it demonstrates the framework for a more restrictive optimization problem than the optimization of an array with a fully excitation driven two-dimensional scan. The reason that this optimization is more restrictive is that the one-dimensional feed network combines static and beam-dependent degrees of freedom, so the optimizer must account for element design together with a partially hard-wired feed structure rather than freely adjusting all excitations for each beam. The purpose of the example is therefore not to cover all possible scan scenarios, but to demonstrate the proposed framework in a setting where both electromagnetic and feed-constraint effects remain visible.
\annotateChangesEnd

\subsection{Electromagnetic Simulation}

For all electromagnetic simulations, our in-house method of moments implementation CMC \cite{noauthor_-house_nodate} is used.

The method of moments uses a grounded dielectric slab \annotateChangesStart on an infinite ground plane\annotateChangesEnd{} \cite{mrochen_characteristic_2025} as Green's function, implying an infinite substrate in the $x$- and $y$\nobreakdash-directions. Edge effects could in principle be included by using a PMCHWT approach \cite{yla-oijala_pmchwt-based_2020} or a windowed Green's function \cite{skrivervik_analysis_1993}. Here, the infinite-substrate assumption is adopted for simplicity because the goal is to demonstrate the proposed optimization method rather than to finalize a particular array design. It also enables a simple and efficient spectral-domain method-of-moments implementation.

\subsection{Initial Considerations}

This section provides initial observations that illustrate the difficulty of the problem and motivate the proposed approach. As a baseline, the array is first designed with a conventional simplified procedure that neglects mutual coupling during the initial design stage. The initial element geometry is shown in Fig.~\ref{fig:initial_element_geometry}~(a). The element is tuned to achieve LHCP in the isolated case, meaning that the edge lengths are chosen such that the relative phase difference between the two orthogonal modes is $90^\circ$, while the port position is chosen such that the amplitudes of the two modes are equal. The isolated element pattern is shown in Fig.~\ref{fig:initial_element_geometry}~(b). The RHCP component is well suppressed in the broadside direction, but the XPR deteriorates away from broadside. For this simple patch geometry, this is already the best XPR achievable in the isolated case.


This initial element is then used to construct the initial array geometry. As a first reference, the array pattern is estimated by superposing the isolated element patterns, as shown in Fig.~\ref{fig:example_array_pattern_element_init_scan}. The row networks are assumed to distribute the power equally across all elements in a row and with equal phase (uniform excitations). In the column network, a Chebyshev taper is used to achieve the desired SLL, and a linear phase progression is used for beam steering. The target SLL and XPR are also indicated in Fig.~\ref{fig:example_array_pattern_element_init_scan}.

The desired SLL is met, whereas the desired XPR is not. More specifically, the result is $\mathrm{XPR}\approx \SI{24}{\decibel}$, consistent with the isolated element pattern. Because all elements are identical, the array pattern can be interpreted as the product of an array factor and the isolated element pattern. The array factor is the same for both polarizations, so the array XPR is determined by the XPR of the isolated element pattern at the target angles. Since the element XPR is already close to the best achievable value for this geometry, identical elements are unlikely to satisfy the desired XPR at both target angles. This indicates that different elements are needed to achieve the desired XPR over the full scan plane.

However, additional challenges arise once mutual coupling is included. The pattern superposition of isolated elements is only valid when coupling is neglected, so a performance drop is expected otherwise. In particular, XPR is expected to degrade because coupling alters the relative phase and amplitude of the two orthogonal modes on each element. This behavior is clearly visible in Fig.~\ref{fig:example_array_pattern_element_init_scan}: when mutual coupling is included and the same excitation coefficients are used in a full-wave simulation, the performance degrades remarkably. This confirms that mutual coupling has a strong impact on array performance and should be considered from the earliest design stage.


\begin{figure}
    \centering

    \definecolor{mycolor1}{rgb}{0.00000,0.44700,0.74100}%
    \definecolor{mycolor2}{rgb}{0.85000,0.32500,0.09800}%
    \definecolor{mycolor3}{rgb}{0.30100,0.74500,0.93300}%
    \definecolor{mycolor4}{rgb}{0.63500,0.07800,0.18400}%

        \input{images/example_array_pattern_element_init_scan_bs.tex}
    \caption{Array pattern of the initial array constructed using a pattern superposition of the isolated element patterns for the target angle $\theta_t = 0^\circ$. The desired LHCP component (\lineleg{mycolor1}) and the undesired RHCP component (\lineleg{mycolor2}) are shown respectively. The desired SLL (\lineleg[dashed]{mycolor3}) and the desired XPR (\lineleg[dashed]{mycolor4}) are denoted by the dashed lines. The dotted lines (\lineleg[dotted]{mycolor1}/\lineleg[dotted]{mycolor2}) show the LHCP and RHCP components of the array pattern when mutual coupling is taken into account and a full-wave simulation of the array is performed.}
    \label{fig:example_array_pattern_element_init_scan}
\end{figure}

\subsection{Preparation of the Proposed Framework}

The proposed optimization framework is now applied to improve the array performance while accounting for mutual coupling from the beginning. The optimization is carried out with Manopt, a toolbox for optimization on manifolds. Before the optimization can start, the inputs to the framework must be specified. The initial geometry was already introduced in the previous subsection. The remaining task is to define the design space by selecting the optimization degrees of freedom and the beams that specify the pattern-to-cost functional.

The formal desired beam definitions $\sigma_s$ in the selected scan range with the desired polarization, SLL and XPR, which are required for the pattern-to-cost functional in \eqref{eq:pattern_to_cost_functional}, are given in Appendix~\ref{sec:app_beams}. 

Next, the design space is defined by selecting the degrees of freedom for the optimization. It is composed as a product manifold of the manifolds of the GSMs of the elements and the manifolds of the excitations.

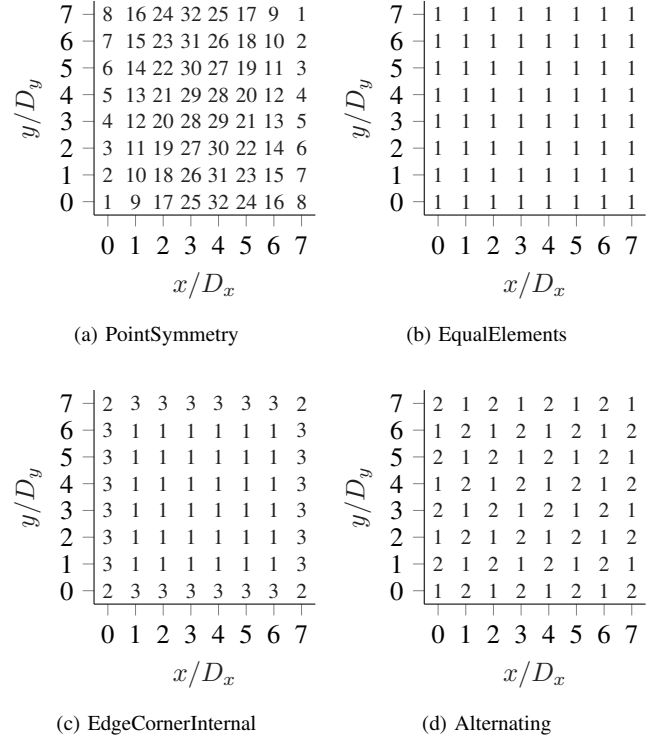
\begin{figure}
    \centering
    \subfloat[PointSymmetry]{
%
%
\definecolor{mycolor1}{rgb}{0.12941,0.12941,0.12941}%
\begin{tikzpicture}

\begin{axis}[%
width=1.1528in,
height=1.115in,
at={(0.2724in,0.3417in)},
scale only axis,
xmin=-0.002676718375,
xmax=0.040150775625,
xtick={0,0.00535343675,0.0107068735,0.01606031025,0.021413747,0.02676718375,0.0321206205,0.03747405725},
xticklabels={{0},{1},{2},{3},{4},{5},{6},{7}},
tick align=outside,
xlabel style={font=\color{mycolor1}},
xlabel={$x/D_x$},
ymin=-0.002676718375,
ymax=0.040150775625,
ytick={0,0.00535343675,0.0107068735,0.01606031025,0.021413747,0.02676718375,0.0321206205,0.03747405725},
yticklabels={{0},{1},{2},{3},{4},{5},{6},{7}},
ylabel style={font=\color{mycolor1}},
ylabel={$y/D_y$},
axis background/.style={fill=white},
axis x line*=bottom,
axis y line*=left,
scaled ticks=false
]
\node[centered, align=center, inner sep=0, font=\color{mycolor1}]
at (axis cs:0,0) {\footnotesize 1};
\node[centered, align=center, inner sep=0, font=\color{mycolor1}]
at (axis cs:0,0.0054) {\footnotesize 2};
\node[centered, align=center, inner sep=0, font=\color{mycolor1}]
at (axis cs:0,0.0107) {\footnotesize 3};
\node[centered, align=center, inner sep=0, font=\color{mycolor1}]
at (axis cs:0,0.0161) {\footnotesize 4};
\node[centered, align=center, inner sep=0, font=\color{mycolor1}]
at (axis cs:0,0.0214) {\footnotesize 5};
\node[centered, align=center, inner sep=0, font=\color{mycolor1}]
at (axis cs:0,0.0268) {\footnotesize 6};
\node[centered, align=center, inner sep=0, font=\color{mycolor1}]
at (axis cs:0,0.0321) {\footnotesize 7};
\node[centered, align=center, inner sep=0, font=\color{mycolor1}]
at (axis cs:0,0.0375) {\footnotesize 8};
\node[centered, align=center, inner sep=0, font=\color{mycolor1}]
at (axis cs:0.0054,0) {\footnotesize 9};
\node[centered, align=center, inner sep=0, font=\color{mycolor1}]
at (axis cs:0.0054,0.0054) {\footnotesize 10};
\node[centered, align=center, inner sep=0, font=\color{mycolor1}]
at (axis cs:0.0054,0.0107) {\footnotesize 11};
\node[centered, align=center, inner sep=0, font=\color{mycolor1}]
at (axis cs:0.0054,0.0161) {\footnotesize 12};
\node[centered, align=center, inner sep=0, font=\color{mycolor1}]
at (axis cs:0.0054,0.0214) {\footnotesize 13};
\node[centered, align=center, inner sep=0, font=\color{mycolor1}]
at (axis cs:0.0054,0.0268) {\footnotesize 14};
\node[centered, align=center, inner sep=0, font=\color{mycolor1}]
at (axis cs:0.0054,0.0321) {\footnotesize 15};
\node[centered, align=center, inner sep=0, font=\color{mycolor1}]
at (axis cs:0.0054,0.0375) {\footnotesize 16};
\node[centered, align=center, inner sep=0, font=\color{mycolor1}]
at (axis cs:0.0107,0) {\footnotesize 17};
\node[centered, align=center, inner sep=0, font=\color{mycolor1}]
at (axis cs:0.0107,0.0054) {\footnotesize 18};
\node[centered, align=center, inner sep=0, font=\color{mycolor1}]
at (axis cs:0.0107,0.0107) {\footnotesize 19};
\node[centered, align=center, inner sep=0, font=\color{mycolor1}]
at (axis cs:0.0107,0.0161) {\footnotesize 20};
\node[centered, align=center, inner sep=0, font=\color{mycolor1}]
at (axis cs:0.0107,0.0214) {\footnotesize 21};
\node[centered, align=center, inner sep=0, font=\color{mycolor1}]
at (axis cs:0.0107,0.0268) {\footnotesize 22};
\node[centered, align=center, inner sep=0, font=\color{mycolor1}]
at (axis cs:0.0107,0.0321) {\footnotesize 23};
\node[centered, align=center, inner sep=0, font=\color{mycolor1}]
at (axis cs:0.0107,0.0375) {\footnotesize 24};
\node[centered, align=center, inner sep=0, font=\color{mycolor1}]
at (axis cs:0.0161,0) {\footnotesize 25};
\node[centered, align=center, inner sep=0, font=\color{mycolor1}]
at (axis cs:0.0161,0.0054) {\footnotesize 26};
\node[centered, align=center, inner sep=0, font=\color{mycolor1}]
at (axis cs:0.0161,0.0107) {\footnotesize 27};
\node[centered, align=center, inner sep=0, font=\color{mycolor1}]
at (axis cs:0.0161,0.0161) {\footnotesize 28};
\node[centered, align=center, inner sep=0, font=\color{mycolor1}]
at (axis cs:0.0161,0.0214) {\footnotesize 29};
\node[centered, align=center, inner sep=0, font=\color{mycolor1}]
at (axis cs:0.0161,0.0268) {\footnotesize 30};
\node[centered, align=center, inner sep=0, font=\color{mycolor1}]
at (axis cs:0.0161,0.0321) {\footnotesize 31};
\node[centered, align=center, inner sep=0, font=\color{mycolor1}]
at (axis cs:0.0161,0.0375) {\footnotesize 32};
\node[centered, align=center, inner sep=0, font=\color{mycolor1}]
at (axis cs:0.0214,0) {\footnotesize 32};
\node[centered, align=center, inner sep=0, font=\color{mycolor1}]
at (axis cs:0.0214,0.0054) {\footnotesize 31};
\node[centered, align=center, inner sep=0, font=\color{mycolor1}]
at (axis cs:0.0214,0.0107) {\footnotesize 30};
\node[centered, align=center, inner sep=0, font=\color{mycolor1}]
at (axis cs:0.0214,0.0161) {\footnotesize 29};
\node[centered, align=center, inner sep=0, font=\color{mycolor1}]
at (axis cs:0.0214,0.0214) {\footnotesize 28};
\node[centered, align=center, inner sep=0, font=\color{mycolor1}]
at (axis cs:0.0214,0.0268) {\footnotesize 27};
\node[centered, align=center, inner sep=0, font=\color{mycolor1}]
at (axis cs:0.0214,0.0321) {\footnotesize 26};
\node[centered, align=center, inner sep=0, font=\color{mycolor1}]
at (axis cs:0.0214,0.0375) {\footnotesize 25};
\node[centered, align=center, inner sep=0, font=\color{mycolor1}]
at (axis cs:0.0268,0) {\footnotesize 24};
\node[centered, align=center, inner sep=0, font=\color{mycolor1}]
at (axis cs:0.0268,0.0054) {\footnotesize 23};
\node[centered, align=center, inner sep=0, font=\color{mycolor1}]
at (axis cs:0.0268,0.0107) {\footnotesize 22};
\node[centered, align=center, inner sep=0, font=\color{mycolor1}]
at (axis cs:0.0268,0.0161) {\footnotesize 21};
\node[centered, align=center, inner sep=0, font=\color{mycolor1}]
at (axis cs:0.0268,0.0214) {\footnotesize 20};
\node[centered, align=center, inner sep=0, font=\color{mycolor1}]
at (axis cs:0.0268,0.0268) {\footnotesize 19};
\node[centered, align=center, inner sep=0, font=\color{mycolor1}]
at (axis cs:0.0268,0.0321) {\footnotesize 18};
\node[centered, align=center, inner sep=0, font=\color{mycolor1}]
at (axis cs:0.0268,0.0375) {\footnotesize 17};
\node[centered, align=center, inner sep=0, font=\color{mycolor1}]
at (axis cs:0.0321,0) {\footnotesize 16};
\node[centered, align=center, inner sep=0, font=\color{mycolor1}]
at (axis cs:0.0321,0.0054) {\footnotesize 15};
\node[centered, align=center, inner sep=0, font=\color{mycolor1}]
at (axis cs:0.0321,0.0107) {\footnotesize 14};
\node[centered, align=center, inner sep=0, font=\color{mycolor1}]
at (axis cs:0.0321,0.0161) {\footnotesize 13};
\node[centered, align=center, inner sep=0, font=\color{mycolor1}]
at (axis cs:0.0321,0.0214) {\footnotesize 12};
\node[centered, align=center, inner sep=0, font=\color{mycolor1}]
at (axis cs:0.0321,0.0268) {\footnotesize 11};
\node[centered, align=center, inner sep=0, font=\color{mycolor1}]
at (axis cs:0.0321,0.0321) {\footnotesize 10};
\node[centered, align=center, inner sep=0, font=\color{mycolor1}]
at (axis cs:0.0321,0.0375) {\footnotesize 9};
\node[centered, align=center, inner sep=0, font=\color{mycolor1}]
at (axis cs:0.0375,0) {\footnotesize 8};
\node[centered, align=center, inner sep=0, font=\color{mycolor1}]
at (axis cs:0.0375,0.0054) {\footnotesize 7};
\node[centered, align=center, inner sep=0, font=\color{mycolor1}]
at (axis cs:0.0375,0.0107) {\footnotesize 6};
\node[centered, align=center, inner sep=0, font=\color{mycolor1}]
at (axis cs:0.0375,0.0161) {\footnotesize 5};
\node[centered, align=center, inner sep=0, font=\color{mycolor1}]
at (axis cs:0.0375,0.0214) {\footnotesize 4};
\node[centered, align=center, inner sep=0, font=\color{mycolor1}]
at (axis cs:0.0375,0.0268) {\footnotesize 3};
\node[centered, align=center, inner sep=0, font=\color{mycolor1}]
at (axis cs:0.0375,0.0321) {\footnotesize 2};
\node[centered, align=center, inner sep=0, font=\color{mycolor1}]
at (axis cs:0.0375,0.0375) {\footnotesize 1};
\end{axis}
\end{tikzpicture}%
    }
    \subfloat[EqualElements]{
%
%
\definecolor{mycolor1}{rgb}{0.12941,0.12941,0.12941}%
\begin{tikzpicture}

\begin{axis}[%
width=1.1528in,
height=1.115in,
at={(0.2724in,0.3417in)},
scale only axis,
xmin=-0.002676718375,
xmax=0.040150775625,
xtick={0,0.00535343675,0.0107068735,0.01606031025,0.021413747,0.02676718375,0.0321206205,0.03747405725},
xticklabels={{0},{1},{2},{3},{4},{5},{6},{7}},
tick align=outside,
xlabel style={font=\color{mycolor1}},
xlabel={$x/D_x$},
ymin=-0.002676718375,
ymax=0.040150775625,
ytick={0,0.00535343675,0.0107068735,0.01606031025,0.021413747,0.02676718375,0.0321206205,0.03747405725},
yticklabels={{0},{1},{2},{3},{4},{5},{6},{7}},
ylabel style={font=\color{mycolor1}},
ylabel={$y/D_y$},
axis background/.style={fill=white},
axis x line*=bottom,
axis y line*=left,
scaled ticks=false
]
\node[centered, align=center, inner sep=0, font=\color{mycolor1}]
at (axis cs:0,0) {\footnotesize 1};
\node[centered, align=center, inner sep=0, font=\color{mycolor1}]
at (axis cs:0,0.0054) {\footnotesize 1};
\node[centered, align=center, inner sep=0, font=\color{mycolor1}]
at (axis cs:0,0.0107) {\footnotesize 1};
\node[centered, align=center, inner sep=0, font=\color{mycolor1}]
at (axis cs:0,0.0161) {\footnotesize 1};
\node[centered, align=center, inner sep=0, font=\color{mycolor1}]
at (axis cs:0,0.0214) {\footnotesize 1};
\node[centered, align=center, inner sep=0, font=\color{mycolor1}]
at (axis cs:0,0.0268) {\footnotesize 1};
\node[centered, align=center, inner sep=0, font=\color{mycolor1}]
at (axis cs:0,0.0321) {\footnotesize 1};
\node[centered, align=center, inner sep=0, font=\color{mycolor1}]
at (axis cs:0,0.0375) {\footnotesize 1};
\node[centered, align=center, inner sep=0, font=\color{mycolor1}]
at (axis cs:0.0054,0) {\footnotesize 1};
\node[centered, align=center, inner sep=0, font=\color{mycolor1}]
at (axis cs:0.0054,0.0054) {\footnotesize 1};
\node[centered, align=center, inner sep=0, font=\color{mycolor1}]
at (axis cs:0.0054,0.0107) {\footnotesize 1};
\node[centered, align=center, inner sep=0, font=\color{mycolor1}]
at (axis cs:0.0054,0.0161) {\footnotesize 1};
\node[centered, align=center, inner sep=0, font=\color{mycolor1}]
at (axis cs:0.0054,0.0214) {\footnotesize 1};
\node[centered, align=center, inner sep=0, font=\color{mycolor1}]
at (axis cs:0.0054,0.0268) {\footnotesize 1};
\node[centered, align=center, inner sep=0, font=\color{mycolor1}]
at (axis cs:0.0054,0.0321) {\footnotesize 1};
\node[centered, align=center, inner sep=0, font=\color{mycolor1}]
at (axis cs:0.0054,0.0375) {\footnotesize 1};
\node[centered, align=center, inner sep=0, font=\color{mycolor1}]
at (axis cs:0.0107,0) {\footnotesize 1};
\node[centered, align=center, inner sep=0, font=\color{mycolor1}]
at (axis cs:0.0107,0.0054) {\footnotesize 1};
\node[centered, align=center, inner sep=0, font=\color{mycolor1}]
at (axis cs:0.0107,0.0107) {\footnotesize 1};
\node[centered, align=center, inner sep=0, font=\color{mycolor1}]
at (axis cs:0.0107,0.0161) {\footnotesize 1};
\node[centered, align=center, inner sep=0, font=\color{mycolor1}]
at (axis cs:0.0107,0.0214) {\footnotesize 1};
\node[centered, align=center, inner sep=0, font=\color{mycolor1}]
at (axis cs:0.0107,0.0268) {\footnotesize 1};
\node[centered, align=center, inner sep=0, font=\color{mycolor1}]
at (axis cs:0.0107,0.0321) {\footnotesize 1};
\node[centered, align=center, inner sep=0, font=\color{mycolor1}]
at (axis cs:0.0107,0.0375) {\footnotesize 1};
\node[centered, align=center, inner sep=0, font=\color{mycolor1}]
at (axis cs:0.0161,0) {\footnotesize 1};
\node[centered, align=center, inner sep=0, font=\color{mycolor1}]
at (axis cs:0.0161,0.0054) {\footnotesize 1};
\node[centered, align=center, inner sep=0, font=\color{mycolor1}]
at (axis cs:0.0161,0.0107) {\footnotesize 1};
\node[centered, align=center, inner sep=0, font=\color{mycolor1}]
at (axis cs:0.0161,0.0161) {\footnotesize 1};
\node[centered, align=center, inner sep=0, font=\color{mycolor1}]
at (axis cs:0.0161,0.0214) {\footnotesize 1};
\node[centered, align=center, inner sep=0, font=\color{mycolor1}]
at (axis cs:0.0161,0.0268) {\footnotesize 1};
\node[centered, align=center, inner sep=0, font=\color{mycolor1}]
at (axis cs:0.0161,0.0321) {\footnotesize 1};
\node[centered, align=center, inner sep=0, font=\color{mycolor1}]
at (axis cs:0.0161,0.0375) {\footnotesize 1};
\node[centered, align=center, inner sep=0, font=\color{mycolor1}]
at (axis cs:0.0214,0) {\footnotesize 1};
\node[centered, align=center, inner sep=0, font=\color{mycolor1}]
at (axis cs:0.0214,0.0054) {\footnotesize 1};
\node[centered, align=center, inner sep=0, font=\color{mycolor1}]
at (axis cs:0.0214,0.0107) {\footnotesize 1};
\node[centered, align=center, inner sep=0, font=\color{mycolor1}]
at (axis cs:0.0214,0.0161) {\footnotesize 1};
\node[centered, align=center, inner sep=0, font=\color{mycolor1}]
at (axis cs:0.0214,0.0214) {\footnotesize 1};
\node[centered, align=center, inner sep=0, font=\color{mycolor1}]
at (axis cs:0.0214,0.0268) {\footnotesize 1};
\node[centered, align=center, inner sep=0, font=\color{mycolor1}]
at (axis cs:0.0214,0.0321) {\footnotesize 1};
\node[centered, align=center, inner sep=0, font=\color{mycolor1}]
at (axis cs:0.0214,0.0375) {\footnotesize 1};
\node[centered, align=center, inner sep=0, font=\color{mycolor1}]
at (axis cs:0.0268,0) {\footnotesize 1};
\node[centered, align=center, inner sep=0, font=\color{mycolor1}]
at (axis cs:0.0268,0.0054) {\footnotesize 1};
\node[centered, align=center, inner sep=0, font=\color{mycolor1}]
at (axis cs:0.0268,0.0107) {\footnotesize 1};
\node[centered, align=center, inner sep=0, font=\color{mycolor1}]
at (axis cs:0.0268,0.0161) {\footnotesize 1};
\node[centered, align=center, inner sep=0, font=\color{mycolor1}]
at (axis cs:0.0268,0.0214) {\footnotesize 1};
\node[centered, align=center, inner sep=0, font=\color{mycolor1}]
at (axis cs:0.0268,0.0268) {\footnotesize 1};
\node[centered, align=center, inner sep=0, font=\color{mycolor1}]
at (axis cs:0.0268,0.0321) {\footnotesize 1};
\node[centered, align=center, inner sep=0, font=\color{mycolor1}]
at (axis cs:0.0268,0.0375) {\footnotesize 1};
\node[centered, align=center, inner sep=0, font=\color{mycolor1}]
at (axis cs:0.0321,0) {\footnotesize 1};
\node[centered, align=center, inner sep=0, font=\color{mycolor1}]
at (axis cs:0.0321,0.0054) {\footnotesize 1};
\node[centered, align=center, inner sep=0, font=\color{mycolor1}]
at (axis cs:0.0321,0.0107) {\footnotesize 1};
\node[centered, align=center, inner sep=0, font=\color{mycolor1}]
at (axis cs:0.0321,0.0161) {\footnotesize 1};
\node[centered, align=center, inner sep=0, font=\color{mycolor1}]
at (axis cs:0.0321,0.0214) {\footnotesize 1};
\node[centered, align=center, inner sep=0, font=\color{mycolor1}]
at (axis cs:0.0321,0.0268) {\footnotesize 1};
\node[centered, align=center, inner sep=0, font=\color{mycolor1}]
at (axis cs:0.0321,0.0321) {\footnotesize 1};
\node[centered, align=center, inner sep=0, font=\color{mycolor1}]
at (axis cs:0.0321,0.0375) {\footnotesize 1};
\node[centered, align=center, inner sep=0, font=\color{mycolor1}]
at (axis cs:0.0375,0) {\footnotesize 1};
\node[centered, align=center, inner sep=0, font=\color{mycolor1}]
at (axis cs:0.0375,0.0054) {\footnotesize 1};
\node[centered, align=center, inner sep=0, font=\color{mycolor1}]
at (axis cs:0.0375,0.0107) {\footnotesize 1};
\node[centered, align=center, inner sep=0, font=\color{mycolor1}]
at (axis cs:0.0375,0.0161) {\footnotesize 1};
\node[centered, align=center, inner sep=0, font=\color{mycolor1}]
at (axis cs:0.0375,0.0214) {\footnotesize 1};
\node[centered, align=center, inner sep=0, font=\color{mycolor1}]
at (axis cs:0.0375,0.0268) {\footnotesize 1};
\node[centered, align=center, inner sep=0, font=\color{mycolor1}]
at (axis cs:0.0375,0.0321) {\footnotesize 1};
\node[centered, align=center, inner sep=0, font=\color{mycolor1}]
at (axis cs:0.0375,0.0375) {\footnotesize 1};
\end{axis}
\end{tikzpicture}%
    } \\[5mm]
    \subfloat[EdgeCornerInternal]{
%
%
\definecolor{mycolor1}{rgb}{0.12941,0.12941,0.12941}%
\begin{tikzpicture}

\begin{axis}[%
width=1.1528in,
height=1.115in,
at={(0.2724in,0.3417in)},
scale only axis,
xmin=-0.002676718375,
xmax=0.040150775625,
xtick={0,0.00535343675,0.0107068735,0.01606031025,0.021413747,0.02676718375,0.0321206205,0.03747405725},
xticklabels={{0},{1},{2},{3},{4},{5},{6},{7}},
tick align=outside,
xlabel style={font=\color{mycolor1}},
xlabel={$x/D_x$},
ymin=-0.002676718375,
ymax=0.040150775625,
ytick={0,0.00535343675,0.0107068735,0.01606031025,0.021413747,0.02676718375,0.0321206205,0.03747405725},
yticklabels={{0},{1},{2},{3},{4},{5},{6},{7}},
ylabel style={font=\color{mycolor1}},
ylabel={$y/D_y$},
axis background/.style={fill=white},
axis x line*=bottom,
axis y line*=left,
scaled ticks=false
]
\node[centered, align=center, inner sep=0, font=\color{mycolor1}]
at (axis cs:0,0) {\footnotesize 2};
\node[centered, align=center, inner sep=0, font=\color{mycolor1}]
at (axis cs:0,0.0054) {\footnotesize 3};
\node[centered, align=center, inner sep=0, font=\color{mycolor1}]
at (axis cs:0,0.0107) {\footnotesize 3};
\node[centered, align=center, inner sep=0, font=\color{mycolor1}]
at (axis cs:0,0.0161) {\footnotesize 3};
\node[centered, align=center, inner sep=0, font=\color{mycolor1}]
at (axis cs:0,0.0214) {\footnotesize 3};
\node[centered, align=center, inner sep=0, font=\color{mycolor1}]
at (axis cs:0,0.0268) {\footnotesize 3};
\node[centered, align=center, inner sep=0, font=\color{mycolor1}]
at (axis cs:0,0.0321) {\footnotesize 3};
\node[centered, align=center, inner sep=0, font=\color{mycolor1}]
at (axis cs:0,0.0375) {\footnotesize 2};
\node[centered, align=center, inner sep=0, font=\color{mycolor1}]
at (axis cs:0.0054,0) {\footnotesize 3};
\node[centered, align=center, inner sep=0, font=\color{mycolor1}]
at (axis cs:0.0054,0.0054) {\footnotesize 1};
\node[centered, align=center, inner sep=0, font=\color{mycolor1}]
at (axis cs:0.0054,0.0107) {\footnotesize 1};
\node[centered, align=center, inner sep=0, font=\color{mycolor1}]
at (axis cs:0.0054,0.0161) {\footnotesize 1};
\node[centered, align=center, inner sep=0, font=\color{mycolor1}]
at (axis cs:0.0054,0.0214) {\footnotesize 1};
\node[centered, align=center, inner sep=0, font=\color{mycolor1}]
at (axis cs:0.0054,0.0268) {\footnotesize 1};
\node[centered, align=center, inner sep=0, font=\color{mycolor1}]
at (axis cs:0.0054,0.0321) {\footnotesize 1};
\node[centered, align=center, inner sep=0, font=\color{mycolor1}]
at (axis cs:0.0054,0.0375) {\footnotesize 3};
\node[centered, align=center, inner sep=0, font=\color{mycolor1}]
at (axis cs:0.0107,0) {\footnotesize 3};
\node[centered, align=center, inner sep=0, font=\color{mycolor1}]
at (axis cs:0.0107,0.0054) {\footnotesize 1};
\node[centered, align=center, inner sep=0, font=\color{mycolor1}]
at (axis cs:0.0107,0.0107) {\footnotesize 1};
\node[centered, align=center, inner sep=0, font=\color{mycolor1}]
at (axis cs:0.0107,0.0161) {\footnotesize 1};
\node[centered, align=center, inner sep=0, font=\color{mycolor1}]
at (axis cs:0.0107,0.0214) {\footnotesize 1};
\node[centered, align=center, inner sep=0, font=\color{mycolor1}]
at (axis cs:0.0107,0.0268) {\footnotesize 1};
\node[centered, align=center, inner sep=0, font=\color{mycolor1}]
at (axis cs:0.0107,0.0321) {\footnotesize 1};
\node[centered, align=center, inner sep=0, font=\color{mycolor1}]
at (axis cs:0.0107,0.0375) {\footnotesize 3};
\node[centered, align=center, inner sep=0, font=\color{mycolor1}]
at (axis cs:0.0161,0) {\footnotesize 3};
\node[centered, align=center, inner sep=0, font=\color{mycolor1}]
at (axis cs:0.0161,0.0054) {\footnotesize 1};
\node[centered, align=center, inner sep=0, font=\color{mycolor1}]
at (axis cs:0.0161,0.0107) {\footnotesize 1};
\node[centered, align=center, inner sep=0, font=\color{mycolor1}]
at (axis cs:0.0161,0.0161) {\footnotesize 1};
\node[centered, align=center, inner sep=0, font=\color{mycolor1}]
at (axis cs:0.0161,0.0214) {\footnotesize 1};
\node[centered, align=center, inner sep=0, font=\color{mycolor1}]
at (axis cs:0.0161,0.0268) {\footnotesize 1};
\node[centered, align=center, inner sep=0, font=\color{mycolor1}]
at (axis cs:0.0161,0.0321) {\footnotesize 1};
\node[centered, align=center, inner sep=0, font=\color{mycolor1}]
at (axis cs:0.0161,0.0375) {\footnotesize 3};
\node[centered, align=center, inner sep=0, font=\color{mycolor1}]
at (axis cs:0.0214,0) {\footnotesize 3};
\node[centered, align=center, inner sep=0, font=\color{mycolor1}]
at (axis cs:0.0214,0.0054) {\footnotesize 1};
\node[centered, align=center, inner sep=0, font=\color{mycolor1}]
at (axis cs:0.0214,0.0107) {\footnotesize 1};
\node[centered, align=center, inner sep=0, font=\color{mycolor1}]
at (axis cs:0.0214,0.0161) {\footnotesize 1};
\node[centered, align=center, inner sep=0, font=\color{mycolor1}]
at (axis cs:0.0214,0.0214) {\footnotesize 1};
\node[centered, align=center, inner sep=0, font=\color{mycolor1}]
at (axis cs:0.0214,0.0268) {\footnotesize 1};
\node[centered, align=center, inner sep=0, font=\color{mycolor1}]
at (axis cs:0.0214,0.0321) {\footnotesize 1};
\node[centered, align=center, inner sep=0, font=\color{mycolor1}]
at (axis cs:0.0214,0.0375) {\footnotesize 3};
\node[centered, align=center, inner sep=0, font=\color{mycolor1}]
at (axis cs:0.0268,0) {\footnotesize 3};
\node[centered, align=center, inner sep=0, font=\color{mycolor1}]
at (axis cs:0.0268,0.0054) {\footnotesize 1};
\node[centered, align=center, inner sep=0, font=\color{mycolor1}]
at (axis cs:0.0268,0.0107) {\footnotesize 1};
\node[centered, align=center, inner sep=0, font=\color{mycolor1}]
at (axis cs:0.0268,0.0161) {\footnotesize 1};
\node[centered, align=center, inner sep=0, font=\color{mycolor1}]
at (axis cs:0.0268,0.0214) {\footnotesize 1};
\node[centered, align=center, inner sep=0, font=\color{mycolor1}]
at (axis cs:0.0268,0.0268) {\footnotesize 1};
\node[centered, align=center, inner sep=0, font=\color{mycolor1}]
at (axis cs:0.0268,0.0321) {\footnotesize 1};
\node[centered, align=center, inner sep=0, font=\color{mycolor1}]
at (axis cs:0.0268,0.0375) {\footnotesize 3};
\node[centered, align=center, inner sep=0, font=\color{mycolor1}]
at (axis cs:0.0321,0) {\footnotesize 3};
\node[centered, align=center, inner sep=0, font=\color{mycolor1}]
at (axis cs:0.0321,0.0054) {\footnotesize 1};
\node[centered, align=center, inner sep=0, font=\color{mycolor1}]
at (axis cs:0.0321,0.0107) {\footnotesize 1};
\node[centered, align=center, inner sep=0, font=\color{mycolor1}]
at (axis cs:0.0321,0.0161) {\footnotesize 1};
\node[centered, align=center, inner sep=0, font=\color{mycolor1}]
at (axis cs:0.0321,0.0214) {\footnotesize 1};
\node[centered, align=center, inner sep=0, font=\color{mycolor1}]
at (axis cs:0.0321,0.0268) {\footnotesize 1};
\node[centered, align=center, inner sep=0, font=\color{mycolor1}]
at (axis cs:0.0321,0.0321) {\footnotesize 1};
\node[centered, align=center, inner sep=0, font=\color{mycolor1}]
at (axis cs:0.0321,0.0375) {\footnotesize 3};
\node[centered, align=center, inner sep=0, font=\color{mycolor1}]
at (axis cs:0.0375,0) {\footnotesize 2};
\node[centered, align=center, inner sep=0, font=\color{mycolor1}]
at (axis cs:0.0375,0.0054) {\footnotesize 3};
\node[centered, align=center, inner sep=0, font=\color{mycolor1}]
at (axis cs:0.0375,0.0107) {\footnotesize 3};
\node[centered, align=center, inner sep=0, font=\color{mycolor1}]
at (axis cs:0.0375,0.0161) {\footnotesize 3};
\node[centered, align=center, inner sep=0, font=\color{mycolor1}]
at (axis cs:0.0375,0.0214) {\footnotesize 3};
\node[centered, align=center, inner sep=0, font=\color{mycolor1}]
at (axis cs:0.0375,0.0268) {\footnotesize 3};
\node[centered, align=center, inner sep=0, font=\color{mycolor1}]
at (axis cs:0.0375,0.0321) {\footnotesize 3};
\node[centered, align=center, inner sep=0, font=\color{mycolor1}]
at (axis cs:0.0375,0.0375) {\footnotesize 2};
\end{axis}
\end{tikzpicture}%
    }
    \subfloat[Alternating]{
%
%
\definecolor{mycolor1}{rgb}{0.12941,0.12941,0.12941}%
\begin{tikzpicture}

\begin{axis}[%
width=1.1528in,
height=1.115in,
at={(0.2724in,0.3417in)},
scale only axis,
xmin=-0.002676718375,
xmax=0.040150775625,
xtick={0,0.00535343675,0.0107068735,0.01606031025,0.021413747,0.02676718375,0.0321206205,0.03747405725},
xticklabels={{0},{1},{2},{3},{4},{5},{6},{7}},
tick align=outside,
xlabel style={font=\color{mycolor1}},
xlabel={$x/D_x$},
ymin=-0.002676718375,
ymax=0.040150775625,
ytick={0,0.00535343675,0.0107068735,0.01606031025,0.021413747,0.02676718375,0.0321206205,0.03747405725},
yticklabels={{0},{1},{2},{3},{4},{5},{6},{7}},
ylabel style={font=\color{mycolor1}},
ylabel={$y/D_y$},
axis background/.style={fill=white},
axis x line*=bottom,
axis y line*=left,
scaled ticks=false
]
\node[centered, align=center, inner sep=0, font=\color{mycolor1}]
at (axis cs:0,0) {\footnotesize 1};
\node[centered, align=center, inner sep=0, font=\color{mycolor1}]
at (axis cs:0,0.0054) {\footnotesize 2};
\node[centered, align=center, inner sep=0, font=\color{mycolor1}]
at (axis cs:0,0.0107) {\footnotesize 1};
\node[centered, align=center, inner sep=0, font=\color{mycolor1}]
at (axis cs:0,0.0161) {\footnotesize 2};
\node[centered, align=center, inner sep=0, font=\color{mycolor1}]
at (axis cs:0,0.0214) {\footnotesize 1};
\node[centered, align=center, inner sep=0, font=\color{mycolor1}]
at (axis cs:0,0.0268) {\footnotesize 2};
\node[centered, align=center, inner sep=0, font=\color{mycolor1}]
at (axis cs:0,0.0321) {\footnotesize 1};
\node[centered, align=center, inner sep=0, font=\color{mycolor1}]
at (axis cs:0,0.0375) {\footnotesize 2};
\node[centered, align=center, inner sep=0, font=\color{mycolor1}]
at (axis cs:0.0054,0) {\footnotesize 2};
\node[centered, align=center, inner sep=0, font=\color{mycolor1}]
at (axis cs:0.0054,0.0054) {\footnotesize 1};
\node[centered, align=center, inner sep=0, font=\color{mycolor1}]
at (axis cs:0.0054,0.0107) {\footnotesize 2};
\node[centered, align=center, inner sep=0, font=\color{mycolor1}]
at (axis cs:0.0054,0.0161) {\footnotesize 1};
\node[centered, align=center, inner sep=0, font=\color{mycolor1}]
at (axis cs:0.0054,0.0214) {\footnotesize 2};
\node[centered, align=center, inner sep=0, font=\color{mycolor1}]
at (axis cs:0.0054,0.0268) {\footnotesize 1};
\node[centered, align=center, inner sep=0, font=\color{mycolor1}]
at (axis cs:0.0054,0.0321) {\footnotesize 2};
\node[centered, align=center, inner sep=0, font=\color{mycolor1}]
at (axis cs:0.0054,0.0375) {\footnotesize 1};
\node[centered, align=center, inner sep=0, font=\color{mycolor1}]
at (axis cs:0.0107,0) {\footnotesize 1};
\node[centered, align=center, inner sep=0, font=\color{mycolor1}]
at (axis cs:0.0107,0.0054) {\footnotesize 2};
\node[centered, align=center, inner sep=0, font=\color{mycolor1}]
at (axis cs:0.0107,0.0107) {\footnotesize 1};
\node[centered, align=center, inner sep=0, font=\color{mycolor1}]
at (axis cs:0.0107,0.0161) {\footnotesize 2};
\node[centered, align=center, inner sep=0, font=\color{mycolor1}]
at (axis cs:0.0107,0.0214) {\footnotesize 1};
\node[centered, align=center, inner sep=0, font=\color{mycolor1}]
at (axis cs:0.0107,0.0268) {\footnotesize 2};
\node[centered, align=center, inner sep=0, font=\color{mycolor1}]
at (axis cs:0.0107,0.0321) {\footnotesize 1};
\node[centered, align=center, inner sep=0, font=\color{mycolor1}]
at (axis cs:0.0107,0.0375) {\footnotesize 2};
\node[centered, align=center, inner sep=0, font=\color{mycolor1}]
at (axis cs:0.0161,0) {\footnotesize 2};
\node[centered, align=center, inner sep=0, font=\color{mycolor1}]
at (axis cs:0.0161,0.0054) {\footnotesize 1};
\node[centered, align=center, inner sep=0, font=\color{mycolor1}]
at (axis cs:0.0161,0.0107) {\footnotesize 2};
\node[centered, align=center, inner sep=0, font=\color{mycolor1}]
at (axis cs:0.0161,0.0161) {\footnotesize 1};
\node[centered, align=center, inner sep=0, font=\color{mycolor1}]
at (axis cs:0.0161,0.0214) {\footnotesize 2};
\node[centered, align=center, inner sep=0, font=\color{mycolor1}]
at (axis cs:0.0161,0.0268) {\footnotesize 1};
\node[centered, align=center, inner sep=0, font=\color{mycolor1}]
at (axis cs:0.0161,0.0321) {\footnotesize 2};
\node[centered, align=center, inner sep=0, font=\color{mycolor1}]
at (axis cs:0.0161,0.0375) {\footnotesize 1};
\node[centered, align=center, inner sep=0, font=\color{mycolor1}]
at (axis cs:0.0214,0) {\footnotesize 1};
\node[centered, align=center, inner sep=0, font=\color{mycolor1}]
at (axis cs:0.0214,0.0054) {\footnotesize 2};
\node[centered, align=center, inner sep=0, font=\color{mycolor1}]
at (axis cs:0.0214,0.0107) {\footnotesize 1};
\node[centered, align=center, inner sep=0, font=\color{mycolor1}]
at (axis cs:0.0214,0.0161) {\footnotesize 2};
\node[centered, align=center, inner sep=0, font=\color{mycolor1}]
at (axis cs:0.0214,0.0214) {\footnotesize 1};
\node[centered, align=center, inner sep=0, font=\color{mycolor1}]
at (axis cs:0.0214,0.0268) {\footnotesize 2};
\node[centered, align=center, inner sep=0, font=\color{mycolor1}]
at (axis cs:0.0214,0.0321) {\footnotesize 1};
\node[centered, align=center, inner sep=0, font=\color{mycolor1}]
at (axis cs:0.0214,0.0375) {\footnotesize 2};
\node[centered, align=center, inner sep=0, font=\color{mycolor1}]
at (axis cs:0.0268,0) {\footnotesize 2};
\node[centered, align=center, inner sep=0, font=\color{mycolor1}]
at (axis cs:0.0268,0.0054) {\footnotesize 1};
\node[centered, align=center, inner sep=0, font=\color{mycolor1}]
at (axis cs:0.0268,0.0107) {\footnotesize 2};
\node[centered, align=center, inner sep=0, font=\color{mycolor1}]
at (axis cs:0.0268,0.0161) {\footnotesize 1};
\node[centered, align=center, inner sep=0, font=\color{mycolor1}]
at (axis cs:0.0268,0.0214) {\footnotesize 2};
\node[centered, align=center, inner sep=0, font=\color{mycolor1}]
at (axis cs:0.0268,0.0268) {\footnotesize 1};
\node[centered, align=center, inner sep=0, font=\color{mycolor1}]
at (axis cs:0.0268,0.0321) {\footnotesize 2};
\node[centered, align=center, inner sep=0, font=\color{mycolor1}]
at (axis cs:0.0268,0.0375) {\footnotesize 1};
\node[centered, align=center, inner sep=0, font=\color{mycolor1}]
at (axis cs:0.0321,0) {\footnotesize 1};
\node[centered, align=center, inner sep=0, font=\color{mycolor1}]
at (axis cs:0.0321,0.0054) {\footnotesize 2};
\node[centered, align=center, inner sep=0, font=\color{mycolor1}]
at (axis cs:0.0321,0.0107) {\footnotesize 1};
\node[centered, align=center, inner sep=0, font=\color{mycolor1}]
at (axis cs:0.0321,0.0161) {\footnotesize 2};
\node[centered, align=center, inner sep=0, font=\color{mycolor1}]
at (axis cs:0.0321,0.0214) {\footnotesize 1};
\node[centered, align=center, inner sep=0, font=\color{mycolor1}]
at (axis cs:0.0321,0.0268) {\footnotesize 2};
\node[centered, align=center, inner sep=0, font=\color{mycolor1}]
at (axis cs:0.0321,0.0321) {\footnotesize 1};
\node[centered, align=center, inner sep=0, font=\color{mycolor1}]
at (axis cs:0.0321,0.0375) {\footnotesize 2};
\node[centered, align=center, inner sep=0, font=\color{mycolor1}]
at (axis cs:0.0375,0) {\footnotesize 2};
\node[centered, align=center, inner sep=0, font=\color{mycolor1}]
at (axis cs:0.0375,0.0054) {\footnotesize 1};
\node[centered, align=center, inner sep=0, font=\color{mycolor1}]
at (axis cs:0.0375,0.0107) {\footnotesize 2};
\node[centered, align=center, inner sep=0, font=\color{mycolor1}]
at (axis cs:0.0375,0.0161) {\footnotesize 1};
\node[centered, align=center, inner sep=0, font=\color{mycolor1}]
at (axis cs:0.0375,0.0214) {\footnotesize 2};
\node[centered, align=center, inner sep=0, font=\color{mycolor1}]
at (axis cs:0.0375,0.0268) {\footnotesize 1};
\node[centered, align=center, inner sep=0, font=\color{mycolor1}]
at (axis cs:0.0375,0.0321) {\footnotesize 2};
\node[centered, align=center, inner sep=0, font=\color{mycolor1}]
at (axis cs:0.0375,0.0375) {\footnotesize 1};
\end{axis}
\end{tikzpicture}%
    }
     \caption{Different strategies for the assignment of the degrees of freedom $\mathbf{\psi}_d$ to the elements. The index $d$ of the degrees of freedom $\mathbf{\psi}_d$ is indicated at the geometric location for all elements $k$ for each strategy.}
     \label{fig:example_dof_assignment}
\end{figure}

The radiation of the elements is described by $N = 2 $ modes and each element is selected to have $P = 1$ port. This means that the GSM of each element is a $3\times 3$ matrix, which is a member of the manifold of unitary symmetric matrices.

For assigning the GSM degrees of freedom to the $K = 64$ elements, different strategies are evaluated. The considered assignment strategies are shown in Fig.~\ref{fig:example_dof_assignment}: PointSymmetry groups point-symmetric element pairs and assigns one degree of freedom per pair; EqualElements assigns one common degree of freedom to all elements; EdgeCornerInternal assigns one degree of freedom each to edge, corner, and internal elements; and Alternating assigns one degree of freedom to each of two alternating element groups.

Next, the manifold of the excitations is defined. Because the feeding network in Fig.~\ref{fig:example_distribution_network} is used, the excitations are partly fixed and partly variable during scanning. In the following, this structure is modeled mathematically. The number of columns of the array is denoted by $C$ and the number of rows by $R$, so that $K = C \cdot R$.

The excitation matrix $\mathbf{v}$ is defined as follows:
\begin{equation}
    \mathbf{v} = \left[ v^{(k)}_s \right]_{k,s} = \mathbf{v}_\mathrm{static} \mathbf{v}_\mathrm{dyn} \in \mathbb{C}^{K \times S},
\end{equation}
with
\begin{equation}
    \mathbf{v}_\mathrm{static} = \begin{bmatrix}
        \mathbf{v}_\mathrm{static}^{1} & ... & \mathbf{0} & \mathbf{0} \\
        \mathbf{0} & \mathbf{v}_\mathrm{static}^{2} & ... & \mathbf{0} \\
        \mathbf{0} & \mathbf{0} & ... & \mathbf{v}_\mathrm{static}^{C}
    \end{bmatrix} \in \mathbb{C}^{K \times C},
\end{equation}
where each $\mathbf{v}_\mathrm{static}^{c}$ is a vector of size
\begin{equation}
    \mathbf{v}_\mathrm{static}^{c} \in \mathbb{C}^{R \times 1},
\end{equation}
which has a norm of:
\begin{equation}
    \|\mathbf{v}_\mathrm{static}^{c}\| = 1, \forall c,
\end{equation}
which is common for all beams.

The dynamic part of the excitation is given by $\mathbf{v}_\mathrm{dyn}$, which is a matrix of size
\begin{equation}
    \mathbf{v}_\mathrm{dyn} \in \mathbb{C}^{C \times S}.
\end{equation}

The corresponding normalization condition is
\begin{equation}
    \sum_{c=1}^{C} |v_\mathrm{dyn}^{c,s}|^2 = 1, \quad \forall s.
\end{equation}

All of these conditions can be expressed as manifold constraints, so the optimization can be carried out with manifold optimization algorithms.

\subsection{Optimization and Initial Results}

\begin{figure}
    \centering
    \input{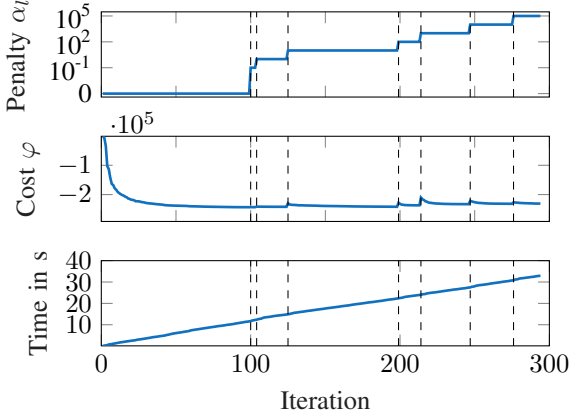}
    \caption{Penalty $\alpha_l$, optimization cost $\varphi$ and total optimization time vs. iteration. The vertical dashed lines denote the iteration indices where the penalty term was increased $\alpha_l \rightarrow \alpha_{l+1}$. }
    \label{fig:example_optim}
\end{figure}

This subsection applies the proposed framework and presents the resulting optimization behavior. The optimization is carried out for strategy~(a), namely the PointSymmetry strategy. The other strategies are discussed and compared with strategy~(a) in the next subsection. Here, the focus is on the optimization process and the results obtained with strategy~(a).

\annotateChangesStart
As a first step, the preprocessing is performed to obtain the modal coupling matrix and the modal far-fields. All preprocessing computations were carried out on a server with two AMD EPYC 7763 CPUs (128 cores / 256 threads in total). For the regular $8\times8$ initial array, the isolated-element quantities had to be computed only once, with the remaining far-fields obtained by phase shifts; this single-core part took \SI{42}{\second}. The modal coupling matrix was then assembled by exploiting the array regularity to reduce the required coupling blocks from $2016$ to $112$ \cite{bleszynski_block-toeplitz_2003}, so that the parallel computation of these terms on $112$ threads required only \SI{38}{\second}. The total preprocessing time was therefore \SI{80}{\second}.
\annotateChangesEnd

The optimization is then carried out with the Manopt toolbox on a single core of a local workstation with an AMD Ryzen AI 9 HX 370 processor. In summary, a single instance of the degrees of freedom $\mathbf{x}$ consists of
\begin{equation}
    \left(N+P\right)^2 \cdot D + C \cdot R + C \cdot S = 3 \cdot 3 \cdot 32 + 8 \cdot 8 + 8 \cdot 13 = 456
\end{equation}
complex double values. The penalty term is increased in every optimization run from $\alpha_1 = 0$, $\alpha_2 = 10^{-1}$, to $\alpha_L = 10^{5}$ in $L = 8$ runs. For every run, the optimization is terminated if the cost function does not decrease by more than $10^{-4}$ for two successive iterations. The \texttt{steepestdescent} algorithm is used within Manopt for the optimization, which is a simple first-order optimization algorithm.

Fig.~\ref{fig:example_optim} shows how the optimization cost $\varphi$ evolves during the optimization process. The cost drops substantially in the first few iterations and then converges to a local minimum. Whenever the penalty term is increased, the cost rises because the optimization problem becomes more restrictive, but it then decreases again and converges to a new local minimum. This pattern repeats until the final penalty term $\alpha_L$ is reached.

The elapsed optimization time per iteration is also shown in Fig.~\ref{fig:example_optim}. It increases almost linearly with the iteration index, and the total optimization time is about $\SI{32}{\second}$ after roughly 290 iterations. For a problem of this size, this short runtime highlights the efficiency of the proposed framework. The main reason is that the optimization acts on the surrogate through GSM and excitation coefficients, which provide a compact abstract representation of the array compared with a full-wave optimization of the geometry. A full-wave optimization with a comparable number of design variables would likely require several hours or days.


\begin{figure}
    \centering
    \input{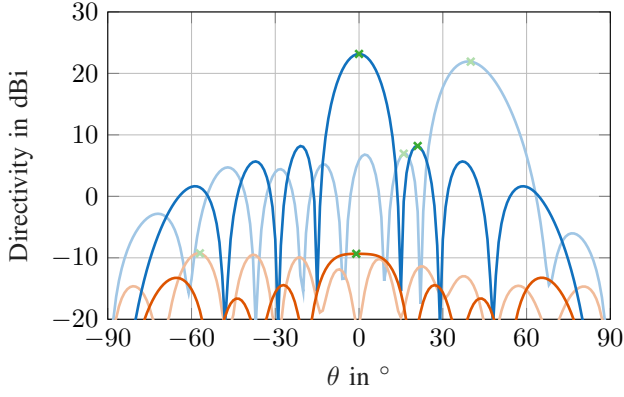}
    \caption{LHCP component (\lineleg{mycolor4}) and RHCP component (\lineleg{mycolor5}) of the directivity of the optimized array for the beams for $\theta_t = 0^\circ$ (solid) and $\theta_t = 40^\circ$ (translucent) for the PointSymmetry strategy. The directivity of the main beam, the maximum sidelobe, and the maximum cross-polarization beam are indicated with markers (\markerleg{mycolor6,mark=x}).}
    \label{fig:example_results_array}
\end{figure}

Fig.~\ref{fig:example_results_array} shows the far-field patterns obtained after optimization. Two representative beams with target angles $\theta_t = 0^\circ$ and $\theta_t = 40^\circ$ are shown. In the surrogate model, both beams satisfy the desired SLL and XPR, demonstrating that the optimization was successful.

\subsection{Evaluating Different Strategies}


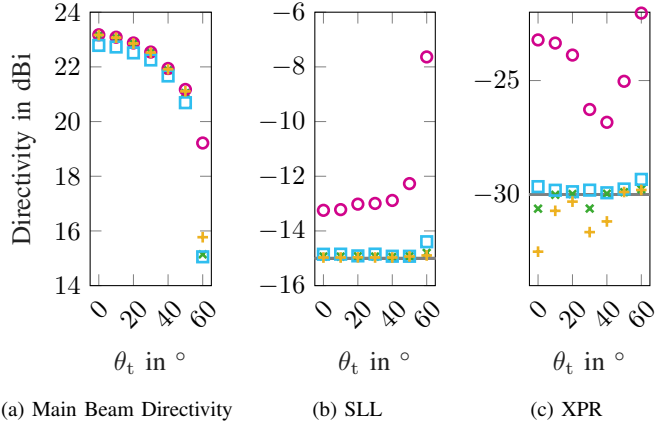
\begin{figure}
    \centering

    \subfloat[Main Beam Directivity]{%
%
%
\definecolor{mycolor1}{rgb}{0.23100,0.66600,0.19600}%
\definecolor{mycolor2}{rgb}{0.81900,0.01500,0.54500}%
\definecolor{mycolor3}{rgb}{0.18400,0.74500,0.93700}%
\definecolor{mycolor4}{rgb}{0.92900,0.69400,0.12500}%
\definecolor{mycolor5}{rgb}{0.12941,0.12941,0.12941}%
\begin{tikzpicture}

\begin{axis}[%
width=0.6314in,
height=1.4241in,
at={(0.3306in,0.3968in)},
scale only axis,
xmin=-5,
xmax=65,
xtick={-60, -40, -20,   0,  20,  40,  60},
xticklabel style={rotate=45},
xlabel style={font=\color{mycolor5}},
xlabel={$\theta_\mathrm{t}$ in $^\circ$},
ymin=14,
ymax=24,
ylabel style={font=\color{mycolor5}},
ylabel={Directivity in dBi},
axis background/.style={fill=white}
]
\addplot [color=mycolor1, line width=1.0pt, only marks, mark=x, mark options={solid, mycolor1}, forget plot]
  table[row sep=crcr]{%
-60	15.143826072963\\
-50	21.0420260069909\\
-40	21.8806015266305\\
-30	22.521783502421\\
-20	22.856382963382\\
-10	23.0616249598595\\
0	23.1490839650346\\
10	23.0616702852639\\
20	22.8563411925613\\
30	22.5217752212134\\
40	21.8797783686355\\
50	21.0419319769294\\
60	15.1442455707064\\
};
\addplot [color=mycolor2, line width=1.0pt, only marks, mark=o, mark options={solid, mycolor2}, forget plot]
  table[row sep=crcr]{%
-60	19.2161031745031\\
-50	21.1745093198466\\
-40	21.941038163647\\
-30	22.5426060373885\\
-20	22.8742757975491\\
-10	23.0850218291698\\
0	23.1739262346929\\
10	23.0850329859515\\
20	22.8742670006147\\
30	22.542597305865\\
40	21.9409254107614\\
50	21.1740600662331\\
60	19.2161318318376\\
};
\addplot [color=mycolor3, line width=1.0pt, only marks, mark=square, mark options={solid, mycolor3}, forget plot]
  table[row sep=crcr]{%
-60	15.0532605448833\\
-50	20.6903880274031\\
-40	21.6680982099069\\
-30	22.2532350383522\\
-20	22.5118126489802\\
-10	22.7271491604528\\
0	22.7852029356254\\
10	22.7271840721887\\
20	22.5117678099095\\
30	22.2531544299998\\
40	21.6681587443277\\
50	20.6902049074559\\
60	15.0538587964669\\
};
\addplot [color=mycolor4, line width=1.0pt, only marks, mark=+, mark options={solid, mycolor4}, forget plot]
  table[row sep=crcr]{%
-60	15.7687384231288\\
-50	21.1070153871902\\
-40	21.9141665459287\\
-30	22.5223236290931\\
-20	22.8557704636198\\
-10	23.0688619653638\\
0	23.165119985956\\
10	23.0687942657974\\
20	22.8555404891844\\
30	22.5223115142557\\
40	21.9141201340947\\
50	21.1064058484226\\
60	15.7710233489938\\
};
\end{axis}

\begin{axis}[%
width=1.063in,
height=1.9685in,
at={(0in,0in)},
scale only axis,
xmin=0,
xmax=1,
ymin=0,
ymax=1,
axis line style={draw=none},
ticks=none,
axis x line*=bottom,
axis y line*=left
]
\end{axis}
\end{tikzpicture}%
    }
    \subfloat[SLL]{
%
%
\definecolor{mycolor1}{rgb}{0.23100,0.66600,0.19600}%
\definecolor{mycolor2}{rgb}{0.81900,0.01500,0.54500}%
\definecolor{mycolor3}{rgb}{0.18400,0.74500,0.93700}%
\definecolor{mycolor4}{rgb}{0.92900,0.69400,0.12500}%
\definecolor{mycolor5}{rgb}{0.12941,0.12941,0.12941}%
\begin{tikzpicture}

\begin{axis}[%
width=0.6275in,
height=1.4241in,
at={(0.192in,0.3968in)},
scale only axis,
xmin=-5,
xmax=65,
xtick={-60, -40, -20,   0,  20,  40,  60},
xticklabel style={rotate=45},
xlabel style={font=\color{mycolor5}},
xlabel={$\theta_\mathrm{t}$ in $^\circ$},
ymin=-16,
ymax=-6,
axis background/.style={fill=white}
]
\addplot [color=mycolor1, line width=1.0pt, only marks, mark=x, mark options={solid, mycolor1}, forget plot]
  table[row sep=crcr]{%
-60	-14.8033248842251\\
-50	-14.9240868809628\\
-40	-14.9505902012102\\
-30	-14.9364511500549\\
-20	-14.924730455612\\
-10	-14.9371870819683\\
0	-14.9350639234256\\
10	-14.937024809301\\
20	-14.9250206435371\\
30	-14.9365760886446\\
40	-14.9508239785134\\
50	-14.9241044284146\\
60	-14.8005361458519\\
};
\addplot [color=gray, forget plot]
  table[row sep=crcr]{%
-5	-15\\
65	-15\\
};
\addplot [color=gray, line width=1.0pt, forget plot]
  table[row sep=crcr]{%
-5	-15\\
65	-15\\
};
\addplot [color=mycolor2, line width=1.0pt, only marks, mark=o, mark options={solid, mycolor2}, forget plot]
  table[row sep=crcr]{%
-60	-7.64164661159384\\
-50	-12.2736783104877\\
-40	-12.8871261007567\\
-30	-12.9953807280349\\
-20	-13.0211795767256\\
-10	-13.2212838386329\\
0	-13.2462412570488\\
10	-13.2202691288992\\
20	-13.0202866347643\\
30	-12.9938641536868\\
40	-12.8849136425726\\
50	-12.2688420229259\\
60	-7.63601712105336\\
};
\addplot [color=gray, forget plot]
  table[row sep=crcr]{%
-5	-15\\
65	-15\\
};
\addplot [color=gray, line width=1.0pt, forget plot]
  table[row sep=crcr]{%
-5	-15\\
65	-15\\
};
\addplot [color=gray, line width=1.0pt, forget plot]
  table[row sep=crcr]{%
-5	-15\\
65	-15\\
};
\addplot [color=mycolor3, line width=1.0pt, only marks, mark=square, mark options={solid, mycolor3}, forget plot]
  table[row sep=crcr]{%
-60	-14.3964525735355\\
-50	-14.9259203931773\\
-40	-14.9250061292788\\
-30	-14.847387151633\\
-20	-14.9126071800383\\
-10	-14.8495271699899\\
0	-14.8519818567243\\
10	-14.8485619064735\\
20	-14.9121022995348\\
30	-14.8471103149746\\
40	-14.9244056162835\\
50	-14.9262092436488\\
60	-14.395224760282\\
};
\addplot [color=gray, forget plot]
  table[row sep=crcr]{%
-5	-15\\
65	-15\\
};
\addplot [color=gray, line width=1.0pt, forget plot]
  table[row sep=crcr]{%
-5	-15\\
65	-15\\
};
\addplot [color=gray, line width=1.0pt, forget plot]
  table[row sep=crcr]{%
-5	-15\\
65	-15\\
};
\addplot [color=gray, line width=1.0pt, forget plot]
  table[row sep=crcr]{%
-5	-15\\
65	-15\\
};
\addplot [color=mycolor4, line width=1.0pt, only marks, mark=+, mark options={solid, mycolor4}, forget plot]
  table[row sep=crcr]{%
-60	-14.8978886211325\\
-50	-14.9361780875496\\
-40	-14.9718226831056\\
-30	-14.9706128189261\\
-20	-14.9654161905774\\
-10	-14.9710235953176\\
0	-14.9721849278189\\
10	-14.9710801629871\\
20	-14.9652291676219\\
30	-14.9703805766961\\
40	-14.9715506256616\\
50	-14.9358597790596\\
60	-14.8963584599591\\
};
\addplot [color=gray, forget plot]
  table[row sep=crcr]{%
-5	-15\\
65	-15\\
};
\end{axis}
\end{tikzpicture}%
    }
    \subfloat[XPR]{%
%
%
\definecolor{mycolor1}{rgb}{0.23100,0.66600,0.19600}%
\definecolor{mycolor2}{rgb}{0.81900,0.01500,0.54500}%
\definecolor{mycolor3}{rgb}{0.18400,0.74500,0.93700}%
\definecolor{mycolor4}{rgb}{0.92900,0.69400,0.12500}%
\definecolor{mycolor5}{rgb}{0.12941,0.12941,0.12941}%
\begin{tikzpicture}

\begin{axis}[%
width=0.6275in,
height=1.4241in,
at={(0.192in,0.3968in)},
scale only axis,
xmin=-5,
xmax=65,
xtick={-60, -40, -20,   0,  20,  40,  60},
xticklabel style={rotate=45},
xlabel style={font=\color{mycolor5}},
xlabel={$\theta_\mathrm{t}$ in $^\circ$},
ymin=-34,
ymax=-22,
axis background/.style={fill=white}
]
\addplot [color=mycolor1, line width=1.0pt, only marks, mark=x, mark options={solid, mycolor1}, forget plot]
  table[row sep=crcr]{%
-60	-29.766674400036\\
-50	-29.8990085397186\\
-40	-29.9687042732455\\
-30	-30.6177874868378\\
-20	-29.9792728281037\\
-10	-30.0135024045041\\
0	-30.6224435838635\\
10	-30.0142288049467\\
20	-29.9802767938666\\
30	-30.6167519490768\\
40	-29.9659499553916\\
50	-29.8987127924051\\
60	-29.7685908300572\\
};
\addplot [color=gray, line width=1.0pt, forget plot]
  table[row sep=crcr]{%
-5	-30\\
65	-30\\
};
\addplot [color=gray, line width=1.0pt, forget plot]
  table[row sep=crcr]{%
-5	-30\\
65	-30\\
};
\addplot [color=mycolor2, line width=1.0pt, only marks, mark=o, mark options={solid, mycolor2}, forget plot]
  table[row sep=crcr]{%
-60	-22.05705698413\\
-50	-25.0579921043335\\
-40	-26.8524694368968\\
-30	-26.2714291201701\\
-20	-23.8783064645937\\
-10	-23.3509342996225\\
0	-23.2165215391324\\
10	-23.3546991941594\\
20	-23.881597690907\\
30	-26.2701021657249\\
40	-26.832024489022\\
50	-25.0329043152009\\
60	-22.0396710842103\\
};
\addplot [color=gray, line width=1.0pt, forget plot]
  table[row sep=crcr]{%
-5	-30\\
65	-30\\
};
\addplot [color=gray, line width=1.0pt, forget plot]
  table[row sep=crcr]{%
-5	-30\\
65	-30\\
};
\addplot [color=gray, line width=1.0pt, forget plot]
  table[row sep=crcr]{%
-5	-30\\
65	-30\\
};
\addplot [color=mycolor3, line width=1.0pt, only marks, mark=square, mark options={solid, mycolor3}, forget plot]
  table[row sep=crcr]{%
-60	-29.3391986125706\\
-50	-29.7599751722962\\
-40	-29.9224095825815\\
-30	-29.8089563881497\\
-20	-29.8789205645537\\
-10	-29.8148024843277\\
0	-29.6551342470338\\
10	-29.8142591767487\\
20	-29.8786662464711\\
30	-29.8091021394232\\
40	-29.9253150452161\\
50	-29.7596686045244\\
60	-29.3369230777711\\
};
\addplot [color=gray, line width=1.0pt, forget plot]
  table[row sep=crcr]{%
-5	-30\\
65	-30\\
};
\addplot [color=gray, line width=1.0pt, forget plot]
  table[row sep=crcr]{%
-5	-30\\
65	-30\\
};
\addplot [color=gray, line width=1.0pt, forget plot]
  table[row sep=crcr]{%
-5	-30\\
65	-30\\
};
\addplot [color=gray, line width=1.0pt, forget plot]
  table[row sep=crcr]{%
-5	-30\\
65	-30\\
};
\addplot [color=mycolor4, line width=1.0pt, only marks, mark=+, mark options={solid, mycolor4}, forget plot]
  table[row sep=crcr]{%
-60	-29.8005414650101\\
-50	-29.8910397332028\\
-40	-31.1873713893273\\
-30	-31.6036798612579\\
-20	-30.3370730695502\\
-10	-30.681709429848\\
0	-32.5095652650091\\
10	-30.7168378844505\\
20	-30.313430106161\\
30	-31.652070366718\\
40	-31.183053705878\\
50	-29.8910997322036\\
60	-29.80142268717\\
};
\addplot [color=gray, line width=1.0pt, forget plot]
  table[row sep=crcr]{%
-5	-30\\
65	-30\\
};
\end{axis}
\end{tikzpicture}%
    }
    \definecolor{mycolor1}{rgb}{0.23100,0.66600,0.19600}%
    \definecolor{mycolor2}{rgb}{0.81900,0.01500,0.54500}
    \definecolor{mycolor3}{rgb}{0.18400,0.74500,0.93700}%
    \definecolor{mycolor4}{rgb}{0.92900,0.69400,0.12500}%
    \definecolor{mycolor5}{rgb}{0.12941,0.12941,0.12941}%
    \caption{Comparison of the achieved high level parameters (a) directivity of the main beam in the desired direction $\theta_t$, (b) sidelobe level and (c) cross-polarization ratio for all beams for the different strategies for the assignment of the degrees of freedom. Strategies: PointSymmetry~(\markerleg{mycolor1,mark=x}), EqualElements~(\markerleg{mycolor2,mark=o}), EdgeCornerInternal~(\markerleg{mycolor3,mark=square}) and Alternating~(\markerleg{mycolor4,mark=+}). The gray lines~(\lineleg{gray}) represent the respective thresholds for the SLL and XPR.}
    \label{fig:example_results_comparison}
\end{figure}

The optimization is then repeated for the different strategies. The modal coupling matrix and the modal far-fields are identical for all strategies because the initial array geometry is unchanged. Consequently, no additional preprocessing is required, which is a significant practical advantage of the proposed framework.

The results are compared in Fig.~\ref{fig:example_results_comparison}. All strategies except EqualElements satisfy the desired SLL and XPR for all beams within the surrogate model. EqualElements does not meet both targets for all beams, although it still performs better than the initial array geometry, whose elements were designed without accounting for mutual coupling.

To complete the procedure, the abstractly optimized results must now be translated into a geometric antenna design, as discussed in the next subsection.

\subsection{Realization}

To realize the array elements, the equations from Section~\ref{sec:realization} are used to determine geometric parameters whose GSMs approximate the optimized targets. As an example, the realization of the first array element ($k = 1$) for the EdgeCornerInternal strategy is shown below. The same procedure is then repeated for all remaining elements.

The first step is to evaluate the optimal free parameter $\chi^{(k)*}$ for the $k$-th element. To this end, the eigenvalue equation is solved for different values of $\chi^{(k)}$ between $0^\circ$ and $180^\circ$. This step is computationally inexpensive because it only involves a $2\times 2$ eigenvalue problem.

\begin{figure}
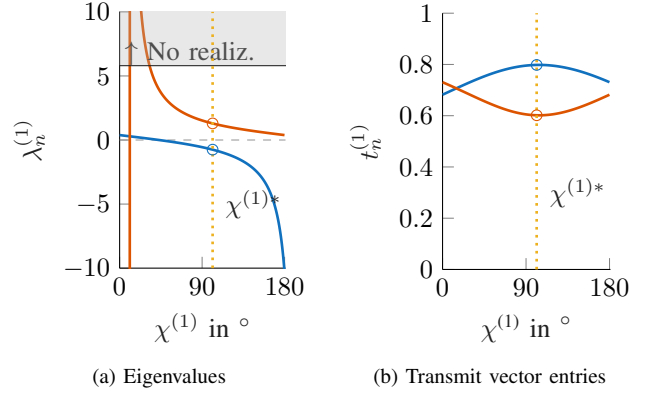

    \centering
    \subfloat[Eigenvalues]{
        \input{images/realization_lambda.tex}
        \label{fig:realization_lambda}
    }
    \subfloat[Transmit vector entries]{
        \input{images/realization_t.tex}
        \label{fig:realization_t}
    }
    \caption{Analysis of the first terminated antenna element for different values of the free parameter $\chi^{(1)}$: (a) Eigenvalues $\lambda_1^{(1)}$ (\lineleg{mycolor1}) and $\lambda_2^{(1)}$ (\lineleg{mycolor2}), and (b) transmit vector entries $t_1^{(1)}$ (\lineleg{mycolor1}) and $t_2^{(1)}$ (\lineleg{mycolor2}). The selected free parameter $\chi^{(1)*}$ is shown as a dashed, vertical line in both subfigures.}
    \label{fig:realization_analysis}
\end{figure}

Fig.~\ref{fig:realization_lambda} shows the resulting eigenvalues $\lambda_1^{(k)}$ and $\lambda_2^{(k)}$ for different values of the free parameter $\chi^{(k)}$ for the first terminated element. The eigenvalues vary substantially with $\chi^{(k)}$. The corresponding transmit-vector entries $t_1^{(k)}$ and $t_2^{(k)}$ in the eigenbasis are shown in Fig.~\ref{fig:realization_t}.

To improve the chances of finding a feasible realization, a low maximum eigenvalue is desirable \cite{manteuffel_compact_2016}. Accordingly, the maximum absolute value of the two eigenvalues is denoted as
\begin{equation}
    \overline{\lambda}^{(k)}(\chi^{(k)})=\max\{|\lambda_1^{(k)}|,|\lambda_2^{(k)}|\},
\end{equation}
and the maximum eigenvalue is minimized by finding the worst parameter $\chi^{(k)}_{\mathrm{max}}$ as follows:
\begin{equation}
    \chi_{\mathrm{max}}^{(k)} = \arg\max_{\chi^{(k)} \in \left[0, 180^\circ\right)} \overline{\lambda}^{(k)}(\chi^{(k)}),
\end{equation}
and adding $90^\circ$, the free parameter $\chi^{(k)*}$ is selected as follows:
\begin{equation}
    \chi^{(k)*} = \chi_{\mathrm{max}}^{(k)} + 90^\circ,
\end{equation}
provided that it is sufficiently far from the poles of the eigenvalues.

\begin{figure}
    \centering
    \subfloat[Element Geometry]{\includegraphics{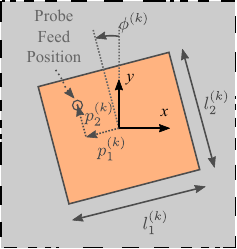}}
    \hspace{5mm}
    \subfloat[Eigenvalue Dependence]{
%
%
\definecolor{mycolor1}{rgb}{0.06600,0.44300,0.74500}%
\definecolor{mycolor2}{rgb}{0.12941,0.12941,0.12941}%
\begin{tikzpicture}

\begin{axis}[%
width=1.0259in,
height=1.0924in,
at={(0.3993in,0.3643in)},
scale only axis,
xmin=0.15,
xmax=0.45,
xlabel style={font=\color{mycolor2}},
xlabel={$l_1^{(k)}/\lambda_0$},
ymin=-11,
ymax=11,
ytick={-10,  -5,   0,   5,  10},
ylabel style={font=\color{mycolor2}},
ylabel={$\lambda_1^{(k)}$},
axis background/.style={fill=white}
]
\addplot [color=mycolor1, line width=1.0pt, forget plot]
  table[row sep=crcr]{%
0.163446406647094	-11.726532193551\\
0.17512114997903	-8.33647507739493\\
0.186795893310965	-5.67072569910913\\
0.1984706366429	-3.55887294444594\\
0.210145379974836	-1.88245149217902\\
0.221820123306771	-0.517509338819746\\
0.233494866638706	0.586297804016625\\
0.245169609970642	1.48725968834424\\
0.256844353302577	2.22856590739535\\
0.268519096634512	2.83883112956883\\
0.280193839966448	3.34913904443531\\
0.291868583298383	3.76988192988443\\
0.303543326630318	4.12372954217018\\
0.315218069962254	4.41804613928831\\
0.326892813294189	4.66633411646028\\
0.338567556626124	4.87389463266018\\
0.35024229995806	5.04714283130905\\
0.361917043289995	5.19409358946404\\
0.37359178662193	5.3167018664575\\
0.385266529953866	5.42037391038244\\
0.396941273285801	5.5084631936863\\
0.408616016617736	5.58096391891634\\
0.420290759949672	5.64222413471053\\
0.431965503281607	5.69308409963567\\
0.443640246613542	5.73540658247348\\
0.455314989945478	5.77039036794035\\
0.466989733277413	5.80057710261228\\
};
\end{axis}
\end{tikzpicture}%
}
    \caption{Geometric realization (a) of the $k$-th antenna element shown together with the dependence of the eigenvalue (b) $\lambda_1^{(k)}$ on the electrical length of the resonant edge for this mode $l_1^{(k)}/\lambda_0$. The dependence of the eigenvalue $\lambda_2^{(k)}$ on the edge length $l_2^{(k)}/\lambda_0$ is similar and is therefore not shown. The rotation angle $\phi^{(k)}$ does not influence the eigenvalues.}
    \label{fig:patch_scattering_altered}
\end{figure}
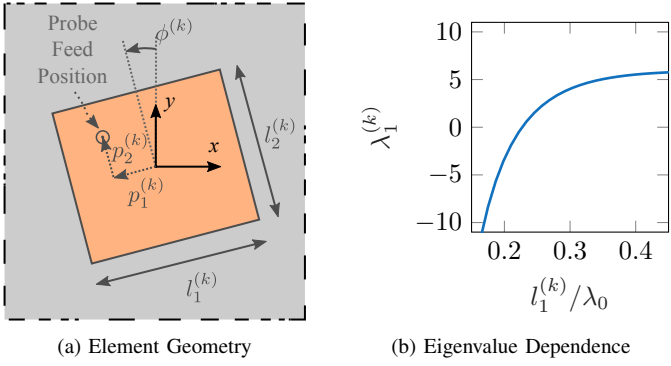

Next, consider the modal transformation matrix $\mathbf{Q}^{(k)}$. It is an orthogonal $2\times2$ matrix and can therefore be interpreted as a rotation matrix. The transformation from the eigenbasis to the original basis can thus be viewed as a rotation in the two-dimensional modal subspace. Because the two modes are the dipole modes of the patch antenna, this modal rotation corresponds to a simple geometric rotation of the patch element about its center, as illustrated in Fig.~\ref{fig:patch_scattering_altered}~(a). The rotation angle $\phi^{(k)}$ of the patch antenna element can then be obtained from $\mathbf{Q}^{(k)}$ as follows:
\begin{equation}
    \phi^{(k)} = -\operatorname{tan}^{-1}\frac{Q_{1,2}^{(k)}}{Q_{1,1}^{(k)}}.
\end{equation}

The edge length $l_1$ of the patch antenna is then varied to determine the dependence of the eigenvalue $\lambda_1$ on $l_1$. The results are shown in Fig.~\ref{fig:patch_scattering_altered}~(b). The eigenvalue $\lambda_1$ changes significantly with $l_1$, so a suitable edge length can be chosen to realize the desired eigenvalue $\lambda_1^{\prime(1)}$. Because the problem is symmetric, $l_1$ mainly influences $\lambda_1$ and $l_2$ mainly influences $\lambda_2$. The same curve can therefore also be used to determine the edge length $l_2$ that realizes the desired eigenvalue $\lambda_2^{\prime(1)}$.

\begin{figure}
    \centering
    \input{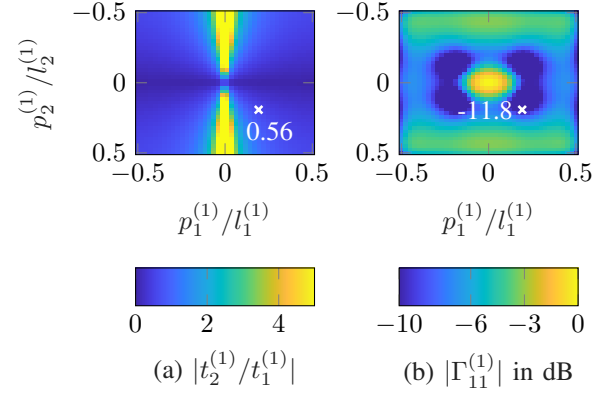}
    \caption{Magnitude of the ratio of the first and second entries of the transmit vector in the eigenbasis $|t_{2}^{(1)}/t_{1}^{(1)}|$ (a) and port input matching $|\Gamma_{11}^{(1)}|$ (b) for varying port positions $(p_1^{(1)}, p_2^{(1)})$ underneath the element. The selected port position $(p_1^{(1)*}, p_2^{(1)*})$ is shown as a white cross.}
    \label{fig:example_port_pos_vary_element_1}
\end{figure}

Finally, the port position is varied to identify a location that realizes the desired transmit vector $\mathbf{t}_{\mathrm{eig}}^{\prime(1)}$ while maintaining good input matching. Fig.~\ref{fig:example_port_pos_vary_element_1} shows the magnitude ratio of the first and second entries of the transmit vector in the eigenbasis, $|t_{2}^{(1)}/t_{1}^{(1)}|$, together with the port input matching $|\Gamma_{11}^{(1)}|$ for varying port positions $(p_1^{(1)}, p_2^{(1)})$. The heatmaps are constructed as described in \cite{yang_computing_2016}.

\begin{figure}
    \centering
%
%
\definecolor{mycolor1}{rgb}{0.34510,0.61961,0.99608}%
\definecolor{mycolor2}{rgb}{1.00000,0.57255,0.23922}%
\definecolor{mycolor3}{rgb}{0.29020,0.77647,0.00000}%
\definecolor{mycolor4}{rgb}{1.00000,0.05098,0.58824}%
\definecolor{mycolor5}{rgb}{1.00000,0.52941,0.83922}%
\definecolor{mycolor6}{rgb}{0.91765,1.00000,0.94510}%
\definecolor{mycolor7}{rgb}{0.12941,0.12941,0.12941}%
\definecolor{mycolor8}{rgb}{0.38824,0.63529,1.00000}%
\definecolor{mycolor9}{rgb}{0.99608,0.56471,0.19216}%
\definecolor{mycolor10}{rgb}{0.46275,0.72157,0.00000}%
\definecolor{mycolor11}{rgb}{0.97255,0.00784,0.58431}%
\definecolor{mycolor12}{rgb}{0.99216,0.58824,0.94510}%
\definecolor{mycolor13}{rgb}{0.98039,0.94902,1.00000}%
\begin{tikzpicture}

\begin{axis}[%
width=0.6204in,
height=0.5797in,
at={(0.4094in,0.7032in)},
scale only axis,
xmin=0.5,
xmax=3.5,
xlabel style={font=\color{mycolor7}},
xlabel={(a) $\widetilde{\mathbf{\Psi}}^{\prime(1)}$},
ymin=0.5,
ymax=3.5,
ytick={1,2,3},
yticklabels={{3},{2},{1}},
axis background/.style={fill=white},
axis x line*=bottom,
axis y line*=left
]

\addplot[area legend, draw=none, fill=mycolor1, forget plot]
table[row sep=crcr] {%
x	y\\
1.5	1.5\\
1.5	0.5\\
0.5	0.5\\
0.5	1.5\\
}--cycle;

\addplot[area legend, draw=none, fill=mycolor2, forget plot]
table[row sep=crcr] {%
x	y\\
1.5	2.5\\
1.5	1.5\\
0.5	1.5\\
0.5	2.5\\
}--cycle;

\addplot[area legend, draw=none, fill=mycolor3, forget plot]
table[row sep=crcr] {%
x	y\\
1.5	3.5\\
1.5	2.5\\
0.5	2.5\\
0.5	3.5\\
}--cycle;

\addplot[area legend, draw=none, fill=mycolor4, forget plot]
table[row sep=crcr] {%
x	y\\
2.5	1.5\\
2.5	0.5\\
1.5	0.5\\
1.5	1.5\\
}--cycle;

\addplot[area legend, draw=none, fill=mycolor5, forget plot]
table[row sep=crcr] {%
x	y\\
2.5	2.5\\
2.5	1.5\\
1.5	1.5\\
1.5	2.5\\
}--cycle;

\addplot[area legend, draw=none, fill=mycolor2, forget plot]
table[row sep=crcr] {%
x	y\\
2.5	3.5\\
2.5	2.5\\
1.5	2.5\\
1.5	3.5\\
}--cycle;

\addplot[area legend, draw=none, fill=mycolor6, forget plot]
table[row sep=crcr] {%
x	y\\
3.5	1.5\\
3.5	0.5\\
2.5	0.5\\
2.5	1.5\\
}--cycle;

\addplot[area legend, draw=none, fill=mycolor4, forget plot]
table[row sep=crcr] {%
x	y\\
3.5	2.5\\
3.5	1.5\\
2.5	1.5\\
2.5	2.5\\
}--cycle;

\addplot[area legend, draw=none, fill=mycolor1, forget plot]
table[row sep=crcr] {%
x	y\\
3.5	3.5\\
3.5	2.5\\
2.5	2.5\\
2.5	3.5\\
}--cycle;
\node[centered, align=center, inner sep=0, font=\color{white!10!black}]
at (axis cs:1,3) {.6};
\node[centered, align=center, inner sep=0, font=\color{white!10!black}]
at (axis cs:1,2) {.5};
\node[centered, align=center, inner sep=0, font=\color{white!10!black}]
at (axis cs:1,1) {.7};
\node[centered, align=center, inner sep=0, font=\color{white!10!black}]
at (axis cs:2,3) {.5};
\node[centered, align=center, inner sep=0, font=\color{white!10!black}]
at (axis cs:2,2) {.5};
\node[centered, align=center, inner sep=0, font=\color{white!10!black}]
at (axis cs:2,1) {.7};
\node[centered, align=center, inner sep=0, font=\color{white!10!black}]
at (axis cs:3,3) {.7};
\node[centered, align=center, inner sep=0, font=\color{white!10!black}]
at (axis cs:3,2) {.7};
\node[centered, align=center, inner sep=0, font=\color{white!10!black}]
at (axis cs:3,1) {.0};
\end{axis}

\begin{axis}[%
width=0.6204in,
height=0.5797in,
at={(1.3197in,0.7032in)},
scale only axis,
xmin=0.5,
xmax=3.5,
xlabel style={font=\color{mycolor7}},
xlabel={(b) $\hat{\mathbf{\Psi}}^{(1)}$},
ymin=0.5,
ymax=3.5,
ytick={1,2,3},
yticklabels={{3},{2},{1}},
axis background/.style={fill=white},
axis x line*=bottom,
axis y line*=left
]

\addplot[area legend, draw=none, fill=mycolor8, forget plot]
table[row sep=crcr] {%
x	y\\
1.5	1.5\\
1.5	0.5\\
0.5	0.5\\
0.5	1.5\\
}--cycle;

\addplot[area legend, draw=none, fill=mycolor9, forget plot]
table[row sep=crcr] {%
x	y\\
1.5	2.5\\
1.5	1.5\\
0.5	1.5\\
0.5	2.5\\
}--cycle;

\addplot[area legend, draw=none, fill=mycolor10, forget plot]
table[row sep=crcr] {%
x	y\\
1.5	3.5\\
1.5	2.5\\
0.5	2.5\\
0.5	3.5\\
}--cycle;

\addplot[area legend, draw=none, fill=mycolor11, forget plot]
table[row sep=crcr] {%
x	y\\
2.5	1.5\\
2.5	0.5\\
1.5	0.5\\
1.5	1.5\\
}--cycle;

\addplot[area legend, draw=none, fill=mycolor12, forget plot]
table[row sep=crcr] {%
x	y\\
2.5	2.5\\
2.5	1.5\\
1.5	1.5\\
1.5	2.5\\
}--cycle;

\addplot[area legend, draw=none, fill=mycolor9, forget plot]
table[row sep=crcr] {%
x	y\\
2.5	3.5\\
2.5	2.5\\
1.5	2.5\\
1.5	3.5\\
}--cycle;

\addplot[area legend, draw=none, fill=mycolor13, forget plot]
table[row sep=crcr] {%
x	y\\
3.5	1.5\\
3.5	0.5\\
2.5	0.5\\
2.5	1.5\\
}--cycle;

\addplot[area legend, draw=none, fill=mycolor11, forget plot]
table[row sep=crcr] {%
x	y\\
3.5	2.5\\
3.5	1.5\\
2.5	1.5\\
2.5	2.5\\
}--cycle;

\addplot[area legend, draw=none, fill=mycolor8, forget plot]
table[row sep=crcr] {%
x	y\\
3.5	3.5\\
3.5	2.5\\
2.5	2.5\\
2.5	3.5\\
}--cycle;
\node[centered, align=center, inner sep=0, font=\color{white!10!black}]
at (axis cs:1,3) {.6};
\node[centered, align=center, inner sep=0, font=\color{white!10!black}]
at (axis cs:1,2) {.5};
\node[centered, align=center, inner sep=0, font=\color{white!10!black}]
at (axis cs:1,1) {.6};
\node[centered, align=center, inner sep=0, font=\color{white!10!black}]
at (axis cs:2,3) {.5};
\node[centered, align=center, inner sep=0, font=\color{white!10!black}]
at (axis cs:2,2) {.4};
\node[centered, align=center, inner sep=0, font=\color{white!10!black}]
at (axis cs:2,1) {.8};
\node[centered, align=center, inner sep=0, font=\color{white!10!black}]
at (axis cs:3,3) {.6};
\node[centered, align=center, inner sep=0, font=\color{white!10!black}]
at (axis cs:3,2) {.8};
\node[centered, align=center, inner sep=0, font=\color{white!10!black}]
at (axis cs:3,1) {.1};
\end{axis}

\begin{axis}[%
width=0.6204in,
height=0.3164in,
at={(2.23in,0.8349in)},
scale only axis,
axis on top,
xmin=0,
xmax=500,
xtick={0,125,250,375,500},
xticklabels={{0},{90},{180},{270},{360}},
xticklabel style={rotate=90},
xlabel style={font=\color{mycolor7}, align=center},
xlabel={$\text{Phase in }^\circ$\\[1ex](c) Colormap},
ymin=0,
ymax=255,
ytick={0,125,250,375,500},
yticklabels={{0},{0.5},{1},{1.5},{2}},
axis background/.style={fill=white},
title style={font=\color{mycolor7}},
title={$\downarrow$ Magnitude}
]
\addplot [forget plot] graphics [xmin=0.5, xmax=500.5, ymin=0.5, ymax=500.5] {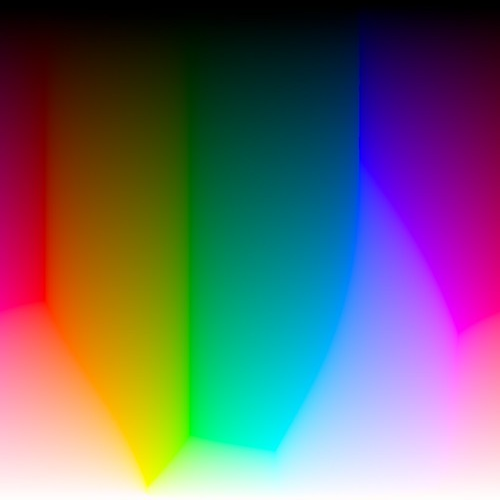};
\end{axis}
\end{tikzpicture}%
    \caption{Comparison of (a) the desired GSM $\widetilde{\mathbf{\Psi}}^{\prime(1)}$ and (b) the realized GSM $\hat{\mathbf{\Psi}}^{(1)}$ of first element of the EdgeCornerInternal array. The colormap that encodes the magnitude and phase is the same for both matrices and is shown in (c).}
    \label{fig:example_element_1_gsm_comp}
\end{figure}
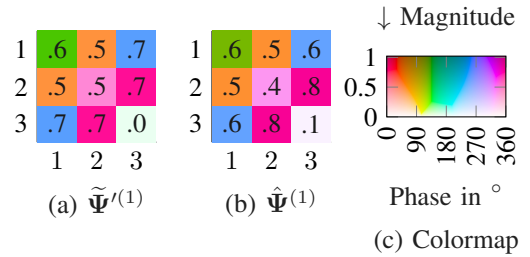

A port position that has a good input matching and realizes the desired transmit vector is selected, which is shown as a white cross in Fig.~\ref{fig:example_port_pos_vary_element_1}. Then, the GSM for this element is calculated by back-transforming the eigenvalues and the transmit vector from the eigenbasis to the original basis according to Appendix~\ref{sec:app_gsm_realized_backtransform}. The resulting GSM is denoted as $\hat{\mathbf{\Psi}}^{(1)}$ and is shown in Fig.~\ref{fig:example_element_1_gsm_comp} together with the desired GSM $\widetilde{\mathbf{\Psi}}^{\prime(1)}$ of the optimized element for comparison.

\begin{figure}
    \centering
    \input{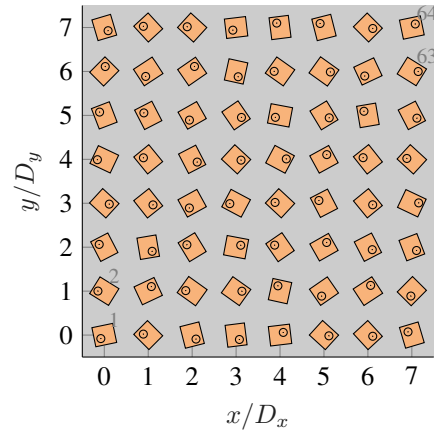}
    \caption{Final array geometry for the PointSymmetry array, which consists of $8\times8$ different patch antenna elements on a grounded dielectric slab with a substrate height of $h=\lambda_0/21$. The global element indices $k$ are shown in grey for the elements 1, 2, 63 and 64 to indicate the ordering. $D_x = D_y = \lambda_0/2$.}
    \label{fig:example_array_modified}
\end{figure}

\annotateChangesStart
The same process is repeated for all elements and all strategies to determine the corresponding geometric parameters. As an example, the final array geometry for the PointSymmetry array is shown in Fig.~\ref{fig:example_array_modified}.

For the PointSymmetry strategy, this post-processing step involves $32$ distinct element realizations. The corresponding computations were distributed between the same EPYC server used for preprocessing and the same local workstation used for optimization. In total, the eigenvalue solutions and near-field calculations for the $32$ selected patch realizations required \SI{589}{\second}, with large parts of this workflow remaining non-parallelized. In addition, the edge-length catalog used for Fig.~\ref{fig:patch_scattering_altered}~(b) was generated on the EPYC server from $27$ parallel runs in \SI{48}{\second}.
\annotateChangesEnd

\subsection{Full-Wave Results and Model Accuracy}

\annotateChangesStart
To verify the results, the realized array geometries are evaluated with full-wave simulations. For this final validation, the full $8\times8$ array impedance matrix has to be assembled with all $2016$ distinct coupling blocks because the realized elements are no longer identical. This full-wave simulation was carried out on the same EPYC server using $256$ parallel threads and required \SI{789}{\second}. Fig.~\ref{fig:example_results_array_fullwave} compares these results with the predictions of the surrogate model. For all three arrays, the predicted and simulated SLL agree well. For the XPR, however, noticeable deviations remain, especially around the main beam, indicating that the surrogate model is not perfectly accurate.
\annotateChangesEnd

Two main effects likely explain these deviations. First, XPR is particularly sensitive to small modeling errors in the mutual-coupling behavior because it depends on a precise phase and amplitude balance between the relevant modes. The surrogate model does not include the current on the probe feed in the modal coupling matrix. Although the direct radiation of the probe feed is more relevant away from broadside, the omitted probe current also perturbs the mutual coupling between elements and thereby the relative modal amplitudes and phases. This affects the predicted XPR, including in the broadside direction. Second, the geometric parameters were realized only on a discrete grid to avoid overly fine-tuned designs and to keep the realized layouts robust. As a result, the realized GSMs deviate from the optimized target GSMs, as illustrated in Fig.~\ref{fig:example_element_1_gsm_comp}.

Despite these limitations, the surrogate model remains sufficiently accurate to guide the design. In particular, the PointSymmetry array shows very good agreement between surrogate and full-wave results. A likely reason is that the realization errors vary from element to element and therefore do not add up coherently in the array pattern. In contrast, for the EdgeCornerInternal and Alternating arrays, similar realization errors occur for many elements, so their effect adds up more coherently and the agreement becomes worse.

If higher accuracy is required in future studies, the final realized array geometry can be reused as the initial geometry for a second optimization run, as in \cite{morlein_deembedding_2025}. This would reduce the mismatch between optimized and realized GSMs and is therefore expected to improve the surrogate accuracy further. It is not pursued here because the present model is already accurate enough to demonstrate successful optimization, which is the main goal of this paper.

\annotateChangesStart
\begin{figure}
\makebox[\linewidth][c]{%
    \hspace*{5mm}%
    \makebox[0.5\linewidth][c]{%
        {%
        \captionsetup[subfloat]{margin={4mm,-4mm}}%
        \subfloat[PointSymmetry, $\theta_{\mathrm{t}} = 0^\circ$]{%
            \input{images/example_results_array_realized_PointSymmetry.tex}%
        }%
        }%
    }%
    \makebox[0.5\linewidth][c]{%
        {%
        \captionsetup[subfloat]{margin={0mm,0mm}}%
        \subfloat[PointSymmetry, $\theta_{\mathrm{t}} = 40^\circ$]{%
            \input{images/example_results_array_realized_PointSymmetry_theta_scan_40.tex}%
        }%
        }%
    }%
}

\vspace{3mm}

\makebox[\linewidth][c]{%
    \hspace*{5mm}%
    \makebox[0.5\linewidth][c]{%
        {%
        \captionsetup[subfloat]{margin={4mm,-4mm}}%
        \subfloat[EdgeCornerInt., $\theta_{\mathrm{t}} = 0^\circ$]{%
            \input{images/example_results_array_realized_EdgeCornerInternal.tex}%
        }%
        }%
    }%
    \makebox[0.5\linewidth][c]{%
        {%
        \captionsetup[subfloat]{margin={0mm,0mm}}%
        \subfloat[EdgeCornerInt., $\theta_{\mathrm{t}} = 40^\circ$]{%
            \input{images/example_results_array_realized_EdgeCornerInternal_theta_scan_40.tex}%
        }%
        }%
    }%
}

\vspace{3mm}

\makebox[\linewidth][c]{%
    \hspace*{5mm}%
    \makebox[0.5\linewidth][c]{%
        {%
        \captionsetup[subfloat]{margin={4mm,-4mm}}%
        \subfloat[Alternating, $\theta_{\mathrm{t}} = 0^\circ$]{%
            \input{images/example_results_array_realized_Alternating.tex}%
        }%
        }%
    }%
    \makebox[0.5\linewidth][c]{%
        {%
        \captionsetup[subfloat]{margin={0mm,0mm}}%
        \subfloat[Alternating, $\theta_{\mathrm{t}} = 40^\circ$]{%
            \input{images/example_results_array_realized_Alternating_theta_scan_40.tex}%
        }%
        }%
    }%
}
    \caption{Comparison of the achieved far-field between the far-field predicted by the surrogate model and the far-field obtained by a full-wave simulation for different strategies for the broadside beam and the beam at $\theta_\mathrm{t} = 40^{\circ}$. The
    LHCP component (\lineleg{mycolor1}) and RHCP component (\lineleg{mycolor2}) of the directivity of the realized array simulated with full-wave. The thick grey curves (\lineleg[line width=2]{white!70!black}) represent the results predicted by the surrogate model. The horizontal solid grey line (\lineleg{white!50!black}) represents the desired XPR.}
    \label{fig:example_results_array_fullwave}
\end{figure}
\annotateChangesEnd

\annotateChangesStart
\subsection{Scalability with Beam-Set Size}

The one-dimensional scan example above demonstrates the proposed framework for a structured feed model with both static and beam-dependent excitation degrees of freedom. Since practical planar arrays often require larger beam sets, a supplementary PointSymmetry study assesses computational scalability for a two-dimensional scan region using beam-dependent variable excitations for each element.

For this scalability study, a deterministic family of two-dimensional beam sets is introduced in which the covered scan sector is enlarged progressively around broadside. The precise definition of these beam sets and the associated two-dimensional constraint regions is given in Appendix~\ref{sec:app_scaling_beams}. In brief, the study increases the number of target beams from a single broadside beam to a square scan sector with $289$ beams covering $[-40^\circ,40^\circ] \times [-40^\circ,40^\circ]$. The other pattern targets are retained, namely $\mathrm{SLL}(\sigma)=\SI{-15}{\dB}$ and $\mathrm{XPR}(\sigma)=\SI{-30}{\dB}$, while the angular sampling used to evaluate the supplementary two-dimensional constraints is reduced to $5^\circ$.

To interpret the computational cost, it is useful to separate array-dependent preprocessing from beam-dependent optimization. Since the initial array configuration remains equal and the modal coupling matrix is independent of the beam set, the previously computed modal coupling matrix can be reused unchanged for the two-dimensional scan study.

The additional cost therefore arises primarily from evaluating the pattern-based objective and its derivatives for a larger number of beams and angular samples. The reduced $5^\circ$ angular sampling changes the discretization of the SLL and XPR constraints, but not the underlying surrogate formulation or beam-set definition.

\begin{figure}
    \centering
%
%
\definecolor{mycolor1}{rgb}{0.06600,0.44300,0.74500}%
\definecolor{mycolor2}{rgb}{0.12941,0.12941,0.12941}%
\begin{tikzpicture}

\begin{axis}[%
width=2.3937in,
height=0.8189in,
at={(0.5984in,2.2047in)},
scale only axis,
bar shift auto,
xmin=0,
xmax=10,
xtick={1,2,3,4,5,6,7,8,9},
xticklabels={\empty},
tick align=outside,
ymin=0,
ymax=0.473134248322148,
ylabel style={font=\color{mycolor2}},
ylabel={Time/Iter. in s},
axis background/.style={fill=white}
]
\addplot[ybar, bar width=0.4, fill=mycolor1, draw=mycolor2, area legend] table[row sep=crcr] {%
1	0.149604283898305\\
2	0.19028646835443\\
3	0.217789050632911\\
4	0.255714616666667\\
5	0.292815666666667\\
6	0.340844796875\\
7	0.358497644444444\\
8	0.418629677083333\\
9	0.473134248322148\\
};
\addplot[forget plot, color=mycolor2] table[row sep=crcr] {%
0	0\\
10	0\\
};
\end{axis}

\begin{axis}[%
width=2.3937in,
height=0.7559in,
at={(0.5984in,1.2913in)},
scale only axis,
bar shift auto,
xmin=0,
xmax=10,
xtick={1,2,3,4,5,6,7,8,9},
xticklabels={\empty},
tick align=outside,
ymin=0,
ymax=447,
ylabel style={font=\color{mycolor2}},
ylabel={Iterations},
axis background/.style={fill=white}
]
\addplot[ybar, bar width=0.4, fill=mycolor1, draw=mycolor2, area legend] table[row sep=crcr] {%
1	236\\
2	79\\
3	79\\
4	60\\
5	57\\
6	64\\
7	90\\
8	96\\
9	447\\
};
\addplot[forget plot, color=mycolor2] table[row sep=crcr] {%
0	0\\
10	0\\
};
\end{axis}

\begin{axis}[%
width=2.3937in,
height=0.6299in,
at={(0.5984in,0.4724in)},
scale only axis,
bar shift auto,
xmin=0,
xmax=10,
xtick={1,2,3,4,5,6,7,8,9},
xticklabels={{$\text{1}^\text{2}$},{$\text{3}^\text{2}$},{$\text{5}^\text{2}$},{$\text{7}^\text{2}$},{$\text{9}^\text{2}$},{$\text{11}^\text{2}$},{$\text{13}^\text{2}$},{$\text{15}^\text{2}$},{$\text{17}^\text{2}$}},
tick align=outside,
xlabel style={font=\color{mycolor2}},
xlabel={Number of Beams},
ymin=0,
ymax=211.567047,
ylabel style={font=\color{mycolor2}},
ylabel={Runtime in s},
axis background/.style={fill=white}
]
\addplot[ybar, bar width=0.4, fill=mycolor1, draw=mycolor2, area legend] table[row sep=crcr] {%
1	35.362574\\
2	15.065743\\
3	17.240971\\
4	15.376932\\
5	16.73545\\
6	21.852773\\
7	32.313757\\
8	40.221846\\
9	211.567047\\
};
\addplot[forget plot, color=mycolor2] table[row sep=crcr] {%
0	0\\
10	0\\
};
\end{axis}
\end{tikzpicture}%
    \caption{Optimization runtime for the supplementary two-dimensional study with progressively enlarged scan sector. The beam sets are constructed from the square grids $\Xi_{\mathrm{t},g}=B_g\times B_g$ with $B_g=\{-5g,-5(g-1),\ldots,0^\circ,\ldots,5g\}$, yielding $(2g+1)^2$ beams. The top plot shows the elapsed time per iteration, the middle plot the number of iterations, and the bottom plot the resulting total optimization time.}
    \label{fig:example_scalability_placeholder}
\end{figure}

Fig.~\ref{fig:example_scalability_placeholder} shows that the elapsed time per iteration increases overall with the number of beams. This is consistent with the larger number of beam-dependent field evaluations entering the SLL and XPR terms, while the modal coupling matrix and modal far-fields are reused unchanged. In contrast, the number of iterations is not monotonic. Very small beam sets constrain only a narrow region around broadside and therefore define a less restrictive optimization problem with a flatter cost landscape, so the optimizer may continue to make small improvements over many iterations. Larger beam sets increase the per-iteration cost, but they also add more off-broadside constraints and thereby guide the optimizer more strongly. Consequently, the total optimization time reflects both effects: a growing beam-dependent evaluation cost per iteration and a non-monotonic convergence behavior across the progressively enlarged scan sectors.

Overall, the results indicate that the optimization remains computationally feasible even for progressively enlarged two-dimensional scan sectors, despite the increasing number of beam-dependent evaluations.

\subsection{Two-Dimensional Pattern Validation}

\begin{figure}
    \centering
    \makebox[\linewidth][c]{%
        \makebox[0.5\linewidth][c]{%
            {%
            \captionsetup[subfloat]{margin={2mm,-8mm},skip=0pt}%
            \subfloat[Surrogate, LHCP]{%
%
%
\definecolor{mycolor1}{rgb}{0.12941,0.12941,0.12941}%
\begin{tikzpicture}

\begin{axis}[%
width=1.1994in,
height=1.0362in,
at={(0.3327in,0.3112in)},
scale only axis,
point meta min=-20,
point meta max=25,
axis on top,
xmin=0,
xmax=360,
xtick={0,90,180,270,360},
xticklabels={{-180},{-90},{0},{90},{180}},
xlabel style={font=\color{mycolor1}},
xlabel={$\phi$},
y dir=reverse,
ymin=0,
ymax=90,
ylabel style={font=\color{mycolor1}},
ylabel={$\theta$},
axis background/.style={fill=white},
axis x line*=bottom,
axis y line*=left
]
\addplot [forget plot] graphics [xmin=-2.5, xmax=357.5, ymin=-0.5, ymax=180.5] {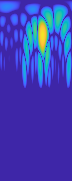};
\end{axis}
\end{tikzpicture}%
%
            }%
            }%
        }%
        \makebox[0.5\linewidth][c]{%
            {%
            \captionsetup[subfloat]{margin={2mm,-8mm},skip=0pt}%
            \subfloat[Full-wave, LHCP]{%
%
%
\definecolor{mycolor1}{rgb}{0.12941,0.12941,0.12941}%
\begin{tikzpicture}

\begin{axis}[%
width=1.1994in,
height=1.0362in,
at={(0.3327in,0.3112in)},
scale only axis,
point meta min=-20,
point meta max=25,
axis on top,
xmin=0,
xmax=360,
xtick={0,90,180,270,360},
xticklabels={{-180},{-90},{0},{90},{180}},
xlabel style={font=\color{mycolor1}},
xlabel={$\phi$},
y dir=reverse,
ymin=0,
ymax=90,
ylabel style={font=\color{mycolor1}},
ylabel={$\theta$},
axis background/.style={fill=white},
axis x line*=bottom,
axis y line*=left
]
\addplot [forget plot] graphics [xmin=-2.5, xmax=357.5, ymin=-0.5, ymax=180.5] {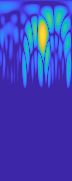};
\end{axis}
\end{tikzpicture}%
%
            }%
            }%
        }%
    }

    \vspace{2mm}

    \makebox[\linewidth][c]{%
        \makebox[0.5\linewidth][c]{%
            {%
            \captionsetup[subfloat]{margin={2mm,-8mm},skip=0pt}%
            \subfloat[Surrogate, RHCP]{%
%
%
\definecolor{mycolor1}{rgb}{0.12941,0.12941,0.12941}%
\begin{tikzpicture}

\begin{axis}[%
width=1.1994in,
height=1.0362in,
at={(0.3327in,0.3112in)},
scale only axis,
point meta min=-20,
point meta max=25,
axis on top,
xmin=0,
xmax=360,
xtick={0,90,180,270,360},
xticklabels={{-180},{-90},{0},{90},{180}},
xlabel style={font=\color{mycolor1}},
xlabel={$\phi$},
y dir=reverse,
ymin=0,
ymax=90,
ylabel style={font=\color{mycolor1}},
ylabel={$\theta$},
axis background/.style={fill=white},
axis x line*=bottom,
axis y line*=left
]
\addplot [forget plot] graphics [xmin=-2.5, xmax=357.5, ymin=-0.5, ymax=180.5] {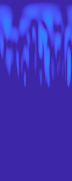};
\end{axis}
\end{tikzpicture}%
%
            }%
            }%
        }%
        \makebox[0.5\linewidth][c]{%
            {%
            \captionsetup[subfloat]{margin={2mm,-8mm},skip=0pt}%
            \subfloat[Full-wave, RHCP]{%
%
%
\definecolor{mycolor1}{rgb}{0.12941,0.12941,0.12941}%
\begin{tikzpicture}

\begin{axis}[%
width=1.1994in,
height=1.0362in,
at={(0.3327in,0.3112in)},
scale only axis,
point meta min=-20,
point meta max=25,
axis on top,
xmin=0,
xmax=360,
xtick={0,90,180,270,360},
xticklabels={{-180},{-90},{0},{90},{180}},
xlabel style={font=\color{mycolor1}},
xlabel={$\phi$},
y dir=reverse,
ymin=0,
ymax=90,
ylabel style={font=\color{mycolor1}},
ylabel={$\theta$},
axis background/.style={fill=white},
axis x line*=bottom,
axis y line*=left
]
\addplot [forget plot] graphics [xmin=-2.5, xmax=357.5, ymin=-0.5, ymax=180.5] {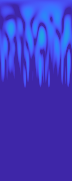};
\end{axis}
\end{tikzpicture}%
%
            }%
            }%
        }%
    }

    \vspace{6mm}

    \makebox[\linewidth][c]{%
        \subfloat[Colorbar]{%
            \input{images/pattern_comp_2d_colorbar.tex}%
        }%
    }
    \caption{Two-dimensional validation for the supplementary PointSymmetry scenario with $13^2$ target beams, shown for the beam $\xi_{\mathrm{t}}=(30^\circ,20^\circ)$. The plots compare the surrogate prediction and the corresponding full-wave result over the full hemisphere for the LHCP and RHCP components.}
    \label{fig:example_2d_validation_placeholder}
\end{figure}

To complement the cut-plane comparisons, Fig.~\ref{fig:example_2d_validation_placeholder} shows a two-dimensional validation result for the supplementary PointSymmetry scenario with $13^2$ target beams. The displayed beam is $\xi_{\mathrm{t}} = (30^\circ,20^\circ)$, and the LHCP and RHCP components are compared between the surrogate prediction and the corresponding full-wave result over the full hemisphere.

The agreement is good overall. In particular, the surrogate reproduces the location and shape of the main beam as well as the dominant sidelobe structure with only moderate deviations. For the undesired RHCP component, both models predict fields of the same order of magnitude over the hemisphere, indicating that the surrogate also captures the cross-polarized response with useful accuracy in this two-dimensional case.

Together with the scalability study, this shows that the model has sufficient accuracy to deliver good optimization results also for two-dimensional scan scenarios over the full hemisphere.
\annotateChangesEnd

\annotateChangesStart

\begin{table*}[!t]
    \centering
    \caption{\annotateChangesStart Comparison with representative coupling-aware array synthesis methods\annotateChangesEnd}
    \label{tab:literature_comparison}
    \footnotesize
    \setlength{\tabcolsep}{4pt}
    \renewcommand{\arraystretch}{1.2}
    \renewcommand{\tabularxcolumn}[1]{m{#1}}
    \begin{tabularx}{\textwidth}{m{0.04\textwidth}m{0.16\textwidth}m{0.08\textwidth}m{0.17\textwidth}m{0.19\textwidth}X}
        \toprule
        \annotateChangesStart \textbf{Work} & \annotateChangesStart \textbf{DoFs} & \annotateChangesStart \textbf{Changing element response?} & \annotateChangesStart \textbf{Coupling model} & \annotateChangesStart \textbf{Reported EM effort} & \annotateChangesStart \textbf{Role of EM simulations} \\
        \midrule
        \annotateChangesStart \cite{yang_circularly_2024} & \annotateChangesStart El. positions, excitations & \annotateChangesStart No & \annotateChangesStart DNN vectorial AEP & \annotateChangesStart 500 FW configurations & \annotateChangesStart Training \\
        \annotateChangesStart \cite{li_machine_2025} & \annotateChangesStart El. positions, excitations & \annotateChangesStart No & \annotateChangesStart GPR/ESA AEP & \annotateChangesStart 90 / 360 ESA configurations & \annotateChangesStart Training \\
        \annotateChangesStart \cite{zhang_physics-informed_2026} & \annotateChangesStart El. positions, excitations & \annotateChangesStart No & \annotateChangesStart DNN AEP & \annotateChangesStart 350 FW configurations & \annotateChangesStart Training \\
        \annotateChangesStart \cite{koziel_rapid_2018} & \annotateChangesStart Excitations & \annotateChangesStart No & \annotateChangesStart Corrected AF & \annotateChangesStart 4--6 FW evaluations & \annotateChangesStart Iterative AF correction \\
        \annotateChangesStart \cite{koziel_expedited_2014} & \annotateChangesStart Excitations & \annotateChangesStart No & \annotateChangesStart Coarse-EM superposition & \annotateChangesStart $\approx 12$ FW evaluations & \annotateChangesStart Surrogate setup and correction \\
        \annotateChangesStart \cite{poli_synthesis_2021} & \annotateChangesStart On/off thinning & \annotateChangesStart No\textsuperscript{a} & \annotateChangesStart Representative EEPs & \annotateChangesStart 5 representative EEPs & \annotateChangesStart EEP library construction \\
        \annotateChangesStart \cite{salmi_optimization_2025} & \annotateChangesStart Reactive loads & \annotateChangesStart Yes: loading & \annotateChangesStart S-matrix and EEPs & \annotateChangesStart 1 FW configuration & \annotateChangesStart Initial EM characterization \\
        \annotateChangesStart \cite{liu_synthesizing_2020} & \annotateChangesStart El. rotation, excitations & \annotateChangesStart Yes: rotation & \annotateChangesStart Transformed AEPs & \annotateChangesStart 4 FW evaluations & \annotateChangesStart Iterative AEP update \\
        \annotateChangesStart \textbf{This work} & \annotateChangesStart \textbf{El. response, excitations} & \annotateChangesStart \textbf{Yes} & \annotateChangesStart \textbf{Modal coupling matrix and MEPs} & \annotateChangesStart \textbf{Impedance matrix assembly and characteristic-mode analysis of initial array}\textsuperscript{b} & \annotateChangesStart \textbf{Modal coupling model construction}\\
        \bottomrule
    \end{tabularx}
    \par\smallskip
    \begin{minipage}{\textwidth}
        \footnotesize
        \annotateChangesStart
        DNN: deep neural network; GPR: Gaussian process regression; ESA: equivalent small array; AEP/EEP/MEP: active/embedded/modal element pattern; AF: array factor; FW: full-wave.
        The effort entries refer to the reported examples and stages, not a common benchmark: training configurations, representative patterns, and FW evaluations are different units of work. Wall-clock times additionally depend on model complexity, solver settings, and hardware.\par
        \textsuperscript{a}Element activation is varied; the individual radiator geometry is not optimized.\quad
        \textsuperscript{b}Element realization and final full-wave validation are separate.
        \annotateChangesEnd
    \end{minipage}
\end{table*}
\section{Comparison With Existing Methods}
\label{sec:literature_comparison}

The following section compares the proposed framework with representative existing coupling-aware synthesis methods. A direct comparison of optimization runtimes or full-wave simulation times across the considered studies is of limited value, because these times are often not reported and depend strongly on model complexity as well as the CPU and solver used. Consequently, differences in reported runtimes cannot be attributed to the synthesis method alone. Since full-wave EM evaluations typically dominate the computational cost of coupling-aware synthesis, their number nevertheless provides a useful indicator of computational effort, although individual evaluations may differ substantially in cost. The comparison therefore focuses on the number and role of full-wave evaluations required for model construction and updating, together with the supported design freedoms, rather than on absolute computation times.

Table~\ref{tab:literature_comparison} summarizes the tradeoff between design freedom and EM effort. Machine-learning-based approaches learn how active element patterns (AEPs) change with the array layout, enabling joint position and excitation optimization \cite{yang_circularly_2024,li_machine_2025,zhang_physics-informed_2026}. This flexibility requires substantial upfront training data: the examples in \cite{yang_circularly_2024,zhang_physics-informed_2026} use 350--500 full-wave configurations. The equivalent small array (ESA) approach in \cite{li_machine_2025} reduces this burden by simulating smaller training structures. Once trained, these models support synthesis without further full-wave evaluations. Their capability to optimization element positions is not included in the present formulation.

For fixed radiators, other methods reduce the cost of obtaining a coupling-aware model for excitation synthesis. The corrected array-factor (AF) approach in \cite{koziel_rapid_2018} avoids a full set of individually simulated embedded element patterns (EEPs), using 4--6 full-wave evaluations to correct the model for amplitude optimization. The coarse-EM superposition approach in \cite{koziel_expedited_2014} instead uses inexpensive coarse-mesh responses and iterative correction, supporting both amplitude and phase optimization. Its tabulated effort is expressed as an equivalent full-wave cost. Poli et al.\ \cite{poli_synthesis_2021} address on/off thinning using a small library of representative EEPs. These approaches economize on EM characterization, but do not optimize the intrinsic response of the radiators.

Closer to the present objective are approaches that allow the element response itself to change. Reactive-load optimization \cite{salmi_optimization_2025} varies terminations on a prescribed structure using an initially characterized S-matrix/EEP model. Liu et al.\ \cite{liu_synthesizing_2020} permit geometric changes through element rotation, jointly optimized with excitation phases. This makes their approach particularly relevant to the present comparison, although the geometric freedom is restricted to rotation. Changes in mutual coupling are captured through successive AEP updates, requiring 4 full-wave evaluations in the reported refinement example.

In contrast, the present framework optimizes element GSMs and excitations with a modal coupling model constructed once from the initial array. This construction requires MoM impedance assembly and characteristic-mode analysis, but no driven full-array MoM solve; subsequent array optimization requires no additional full-wave evaluations. Thus, response changes beyond prescribed loading or rotation can be explored without either a training dataset or iterative full-wave model updates. The scope remains limited by the retained modal basis and the lossless reciprocal GSM parameterization, and physical realization and final validation require separate EM work. The comparison therefore highlights the EM requirements associated with different design freedoms, rather than inferring runtime speedups from unlike simulation counts.
\annotateChangesEnd

\section{Conclusion}
\label{sec:conclusion}

A computationally efficient framework for coupling-aware antenna-array synthesis based on characteristic modes and generalized scattering matrices was presented. By separating electromagnetic preprocessing from optimization variables, the method enables rapid iterative design while preserving a physically grounded treatment of mutual coupling. Modeling element behavior on the manifold of unitary symmetric matrices naturally enforces reciprocity and losslessness.

The $8\times8$ circularly polarized phased array example showed that the workflow can jointly optimize element GSMs and excitations for multiple beams under sidelobe and cross-polarization constraints. A repeated-penalty strategy effectively guided the optimizer from gain-oriented to constraint-compliant solutions. The optimization converged in a short runtime on a single CPU core, and full-wave results confirmed the main surrogate trends.
In particular, design strategies with largely non-identical element assignments achieved the required SLL and yielded strong XPR performance.

\annotateChangesStart
An additional scalability study showed that the optimization time increases with the number of target beams, while larger two-dimensional beam sets remain practically manageable within the proposed framework.

Overall, the results demonstrate that the framework is flexible enough to accommodate different design strategies and can be extended to more complex scenarios, making it a promising tool for efficient antenna array synthesis under the consideration of mutual coupling.
\annotateChangesEnd

\section*{Acknowledgment}

During the preparation of this work, the authors used the AI tools ChatGPT and DeepL to improve formulation and wording, while the scientific content and the initial text drafts were developed entirely by the human authors. After using these tools, the authors reviewed and edited the content as needed and take full responsibility for the published article.

\appendices

\section{Beam Definitions for the Example Array}
\label{sec:app_beams}

\begin{table}[!tb]
\centering
\caption{Multi-beam optimization parameters}
\label{tab:multibeam_params}
\renewcommand{\arraystretch}{2}
\begin{tabular}{ccrccc}
\toprule
Beam & $\xi_{\mathrm{t}}(\sigma)$ & $M_{\mathrm{S}}(\sigma)$ & $M_{\mathrm{X}}(\sigma)$ & $\mathrm{SLL}(\sigma)$ & $\mathrm{XPR}(\sigma)$ \\
\midrule
$\sigma_1$ & $(-60^\circ, 0^\circ)$ & $\prescript{-90^\circ}{-25^\circ}{\mathlarger{\mathlarger{\mathlarger{\Xi}}}}$ & \multirow{13}{*}{$\mathlarger{\mathlarger{\mathlarger{\Xi_0}}}$} & \multirow{13}{*}{-15dB} & \multirow{13}{*}{-30dB} \\\cmidrule(lr){1-3}
$\sigma_2$ & $(-50^\circ, 0^\circ)$ & $\prescript{-75^\circ}{-30^\circ}{\mathlarger{\mathlarger{\mathlarger{\Xi}}}}$ & & & \\\cmidrule(lr){1-3}
$\sigma_3$ & $(-40^\circ, 0^\circ)$ & $\prescript{-60^\circ}{-25^\circ}{\mathlarger{\mathlarger{\mathlarger{\Xi}}}}$ & & & \\\cmidrule(lr){1-3}
$\sigma_4$ & $(-30^\circ, 0^\circ)$ & $\prescript{-50^\circ}{-15^\circ}{\mathlarger{\mathlarger{\mathlarger{\Xi}}}}$ & & & \\\cmidrule(lr){1-3}
$\sigma_5$ & $(-20^\circ, 0^\circ)$ & $\prescript{-35^\circ}{-5^\circ}{\mathlarger{\mathlarger{\mathlarger{\Xi}}}}$ & & & \\\cmidrule(lr){1-3}
$\sigma_6$ & $(-10^\circ, 0^\circ)$ & $\prescript{-25^\circ}{5^\circ}{\mathlarger{\mathlarger{\mathlarger{\Xi}}}}$ & & & \\\cmidrule(lr){1-3}
$\sigma_7$ & $(0^\circ, 0^\circ)$ & $\prescript{-15^\circ}{15^\circ}{\mathlarger{\mathlarger{\mathlarger{\Xi}}}}$ & & & \\\cmidrule(lr){1-3}
$\sigma_8$ & $(10^\circ, 0^\circ)$ & $\prescript{-5^\circ}{25^\circ}{\mathlarger{\mathlarger{\mathlarger{\Xi}}}}$ & & & \\\cmidrule(lr){1-3}
$\sigma_9$ & $(20^\circ, 0^\circ)$ & $\prescript{5^\circ}{35^\circ}{\mathlarger{\mathlarger{\mathlarger{\Xi}}}}$ & & & \\\cmidrule(lr){1-3}
$\sigma_{10}$ & $(30^\circ, 0^\circ)$ & $\prescript{15^\circ}{50^\circ}{\mathlarger{\mathlarger{\mathlarger{\Xi}}}}$ & & & \\\cmidrule(lr){1-3}
$\sigma_{11}$ & $(40^\circ, 0^\circ)$ & $\prescript{25^\circ}{60^\circ}{\mathlarger{\mathlarger{\mathlarger{\Xi}}}}$ & & & \\\cmidrule(lr){1-3}
$\sigma_{12}$ & $(50^\circ, 0^\circ)$ & $\prescript{30^\circ}{75^\circ}{\mathlarger{\mathlarger{\mathlarger{\Xi}}}}$ & & & \\\cmidrule(lr){1-3}
$\sigma_{13}$ & $(60^\circ, 0^\circ)$ & $\prescript{25^\circ}{90^\circ}{\mathlarger{\mathlarger{\mathlarger{\Xi}}}}$ & & & \\
\bottomrule
\end{tabular}
\end{table}

The beam definitions $\sigma_s$, which specify the target angles, the desired SLL and the desired XPR in \eqref{eq:pattern_to_cost_functional} are given in Table~\ref{tab:multibeam_params}.

Since the array is designed to scan in a single plane at $\phi = 0^\circ$, the target angles are defined as $\theta_t \in \{-60^\circ, -50^\circ, ..., 50^\circ, 60^\circ\}$ and $\phi_t = 0^\circ$ for all beams. The desired SLL is defined as $\mathrm{SLL} = \SI{-15}{\dB}$ and the desired XPR is defined as $\mathrm{XPR} = \SI{-30}{\dB}$ for all beams.

For the specification of the angular regions of the sidelobe in the radiation pattern, the discrete angular set is defined as:
\begin{equation}
    \begin{aligned}
    \prescript{\theta_A}{\theta_B}{\Xi} = \Big\{(\theta, 0^\circ) \Big| \, \theta \in \mathbb{Z},\, & (-90^\circ \leq \theta \leq \theta_A) \\ &
    \vee (\theta_B \leq \theta \leq 90^\circ) \Big\},
    \end{aligned}
\end{equation}
where $\theta_A$ and $\theta_B$ denote the lower and upper bounds of the main beam range in degrees, respectively. This set contains all integer degree angles outside of the specified range. For example, for the beam $\sigma_2$, the sidelobe region is defined as:
\begin{equation}
    \begin{aligned}
    M_{\mathrm{S}}(\sigma_2) = \prescript{-70^\circ}{-35^\circ}{\Xi} = \{&(-90^\circ, 0^\circ), (-89^\circ, 0^\circ), \ldots, \\ &(-70^\circ, 0^\circ), (-35^\circ, 0^\circ), \ldots, \\ &(90^\circ, 0^\circ)\}
    \end{aligned}
\end{equation}

Furthermore, for the cross-polarization region, the set is defined as:
\begin{equation}
    \Xi_0 = \Big\{(\theta, 0^\circ) \Big| \, \theta \in \mathbb{Z}, (-90^\circ \leq \theta \leq 90^\circ) \Big\},
\end{equation}

\annotateChangesStart

\section{Supplementary Beam Sets for the Scaling Study}
\label{sec:app_scaling_beams}

For the supplementary scalability and two-dimensional validation study discussed in Section~\ref{sec:example}, a separate family of progressively enlarged two-dimensional beam sets is used. For an integer parameter $g \in \{0,1,\ldots,8\}$, define
\begin{equation}
    B_g = \{-5g,-5(g-1),\ldots,0^\circ,\ldots,5g\},
\end{equation}
\begin{equation}
    \Xi_{\mathrm{t},g} = B_g \times B_g.
\end{equation}
This yields
\begin{equation}
    |\Xi_{\mathrm{t},g}| = (2g+1)^2 \in \{1,9,25,49,81,121,169,225,289\}
\end{equation}
for $g=0,1,\ldots,8$, respectively. In this way, the scan sector is enlarged deterministically from the broadside beam to the square sector $[-40^\circ,40^\circ] \times [-40^\circ,40^\circ]$, while the angular spacing between neighboring target directions remains fixed at $5^\circ$. For all beams of this supplementary study, the same target values $\mathrm{SLL}(\sigma)=\SI{-15}{\dB}$ and $\mathrm{XPR}(\sigma)=\SI{-30}{\dB}$ are used.

To keep the computational and memory cost manageable for the enlarged beam sets, the angular sampling used to define the supplementary two-dimensional constraint sets is reduced to steps of $5^\circ$. For the corresponding two-dimensional angular regions, let
\begin{equation}
    \begin{aligned}
        \Xi_{\mathrm{2D}} = \bigl\{(\theta,\phi) \bigm|\;& \theta,\phi \in 5\mathbb{Z}, \\
        & -90^\circ \leq \theta \leq 90^\circ, \\
        & -90^\circ \leq \phi \leq 90^\circ \bigr\}
    \end{aligned}
\end{equation}
denote the discrete angular evaluation domain. Furthermore, write the target direction of beam $\sigma$ as $\xi_{\mathrm{t}}(\sigma) = (\theta_{\mathrm{t}}(\sigma),\phi_{\mathrm{t}}(\sigma))$. Then a simple two-dimensional analogue of the sidelobe region is
\begin{equation}
    \begin{aligned}
        M_{\mathrm{S},\mathrm{2D}}(\sigma) = \bigl\{(\theta,\phi) \in \Xi_{\mathrm{2D}} \bigm|\;&
        |\theta-\theta_{\mathrm{t}}(\sigma)| > 15^\circ \\
        &{}\vee |\phi-\phi_{\mathrm{t}}(\sigma)| > 15^\circ \bigr\},
    \end{aligned}
\end{equation}
which excludes a rectangular main-beam window of width $30^\circ$ around the target direction. Analogously, the cross-polarization region is defined as
\begin{equation}
    M_{\mathrm{X},\mathrm{2D}}(\sigma) = \Xi_{\mathrm{2D}},
\end{equation}
so that the XPR requirement is enforced on the complete sampled angular domain.

\annotateChangesEnd

\section{Relation to the Terminated Structure}
\label{sec:app_terminated}

This appendix collects the derivation that links the GSM of the realized antenna with ports to the method-of-moments description of the corresponding terminated structure. The material is used for realization, but it is not required to understand the optimization workflow in the main text.

\subsection{Terminated Structure and Modal Transformation Matrix}
\label{sec:def_modified_array}

For the realization of the optimized GSMs of the elements, the problem can be formulated as follows: Given the desired modal transformation matrix $\widetilde{\mathbf{Q}}^{(k)}$, the desired eigenvalues $\widetilde{\lambda}_n^{(k)}$, the desired transmit matrix $\widetilde{\mathbf{T}}_{\mathrm{eig}}^{(k)}$ in the eigenbasis and the desired port-input matching $\widetilde{\mathbf{\Gamma}}^{(k)}$, find the geometric parameters of the antenna element that realizes these desired parameters.

This problem can be addressed through repeated method-of-moments analyses of modified element-level structures, in which the geometric realization parameters are varied to find a good match.

For each element, the isolated self-impedance matrix $\hat{\mathbf{Z}}^{(k,k)}$ is assembled:
\begin{equation}
    \hat{\mathbf{Z}}^{(k,k)} \hat{\mathbf{I}}^{(k)}_{\mathrm{iso}} = \hat{\mathbf{V}}^{(k)}_{\mathrm{iso}}.
\end{equation}

The characteristic modes of this terminated structure satisfy
\begin{equation}
    \operatorname{Im} \hat{\mathbf{Z}}^{(k,k)} \hat{\mathbf{I}}_{\mathrm{CM},n}^{(k)} = \hat{\lambda}_n^{(k)} \operatorname{Re} \hat{\mathbf{Z}}^{(k,k)} \hat{\mathbf{I}}_{\mathrm{CM},n}^{(k)},
\end{equation}
which yields the eigenbasis scattering matrix
\begin{equation}
    \hat{\mathbf{S}}_{0,\mathrm{eig}}^{(k)} = -\operatorname{diag} \frac{1 - j \hat{\lambda}_n^{(k)}}{1 + j \hat{\lambda}_n^{(k)}}.
\end{equation}

To express this terminated structure in the common modal basis of the initial array, define the modal transformation matrix
\begin{equation}
    \hat{\mathbf{Q}}^{(k)} = \mathbf{I}_{\mathrm{CM}}^{(k)\mathrm{T}} \hat{\mathbf{R}}^{(k,k)} \hat{\mathbf{I}}_{\mathrm{CM}}^{(k)},
\end{equation}
where the cross-radiation matrix is
\begin{equation}
    \hat{\mathbf{R}}^{(k,k)} = \left[ \int_{\hat{\Omega}_k} \int_{\Omega_k} \mathbf{\psi}_\mu^{(k)}(\mathbf{r}) \operatorname{Re} \mathbf{G}(\mathbf{r},\mathbf{r}') \hat{\mathbf{\psi}}_m^{(k)}(\mathbf{r}') dS dS' \right]_{\mu m},
\end{equation}
where $\hat{\Omega}_k$ is the surface of the terminated structure, $\Omega_k$ is the surface of the initial element geometry, and $\psi_\mu^{(k)}$ and $\hat{\psi}_m^{(k)}$ are the basis functions of the terminated structure and the initial element geometry, respectively.

\begin{equation}
    \hat{\mathbf{T}}^{(k)}_\mathrm{eig} = \left[\hat{t}^{(k)}_{n,p}\right]_{n,p} = \left[ \frac{1}{1 + \mathrm{j} \hat{\lambda}_n^{(k)}} \hat{\mathbf{I}}_{\mathrm{CM},n}^{(k)\mathrm{T}} \hat{\mathbf{Z}}^{(k,k)} \hat{\mathbf{I}}_p^{(k)} \right]_{n,p},
\end{equation}
where $\hat{\mathbf{I}}_p^{(k)}$ is the current distribution obtained when the $p$\nobreakdash-th port of the isolated $k$\nobreakdash-th element is excited with unit amplitude $v_{p}^{(k)}= 1$.

The structure $\hat{\Omega}$ can then be modified iteratively until the desired parameters $\widetilde{\mathbf{Q}}^{(k)}$, $\widetilde{\lambda}_n^{(k)}$, $\widetilde{\mathbf{T}}_{\mathrm{eig}}^{(k)}$, and $\widetilde{\mathbf{\Gamma}}^{(k)}$ are achieved, with $\hat{\mathbf{\Gamma}}^{(k)}$ calculated according to the standard method-of-moments port implementation.

\subsection{GSM of the Realized Antenna in the Initial Basis}
\label{sec:app_gsm_realized_backtransform}

Let
\begin{equation}
        \hat{\mathbf{\Psi}}^{(k)}_\mathrm{eig}
        = \begin{bmatrix}
            \hat{\mathbf{S}}^{(k)}_\mathrm{eig} & \hat{\mathbf{T}}^{(k)}_\mathrm{eig} \\
            \hat{\mathbf{R}}^{(k)}_\mathrm{eig} & \hat{\mathbf{\Gamma}}^{(k)}
        \end{bmatrix}
\end{equation}
with
\begin{equation}
    \hat{\mathbf{S}}^{(k)}_\mathrm{eig} = \hat{\mathbf{S}}_{0,\mathrm{eig}}^{(k)} - \hat{\mathbf{T}}^{(k)}_\mathrm{eig} \left( \mathbf{\Gamma}_\mathrm{L} - \hat{\mathbf{\Gamma}}^{(k)} \right)^{-1} \hat{\mathbf{R}}^{(k)}_\mathrm{eig},
\end{equation}
be the GSM of the realized antenna in the eigenbasis of the terminated antenna. Then, the GSM of the realized antenna with ports in the common modal basis of the initial antenna is given by
\begin{equation}
    \hat{\mathbf{\Psi}}^{(k)} = \begin{bmatrix}
        \mathbf{I} + \hat{\mathbf{Q}}^{(k)} \left(\hat{\mathbf{S}}^{(k)}_\mathrm{eig} - \mathbf{I} \right) \hat{\mathbf{Q}}^{(k)\mathrm{T}} & \hat{\mathbf{Q}}^{(k)} \hat{\mathbf{T}}^{(k)}_\mathrm{eig} \\
        \hat{\mathbf{R}}^{(k)}_\mathrm{eig} \hat{\mathbf{Q}}^{(k)\mathrm{T}} & \hat{\mathbf{\Gamma}}^{(k)}
    \end{bmatrix}.
\end{equation}








\ifCLASSOPTIONcaptionsoff
  \newpage
\fi



\bibliographystyle{IEEEtran}
\bibliography{IEEEabrv,lit}

\begin{IEEEbiography}[{\includegraphics[width=1in,height=1.25in,clip,keepaspectratio]{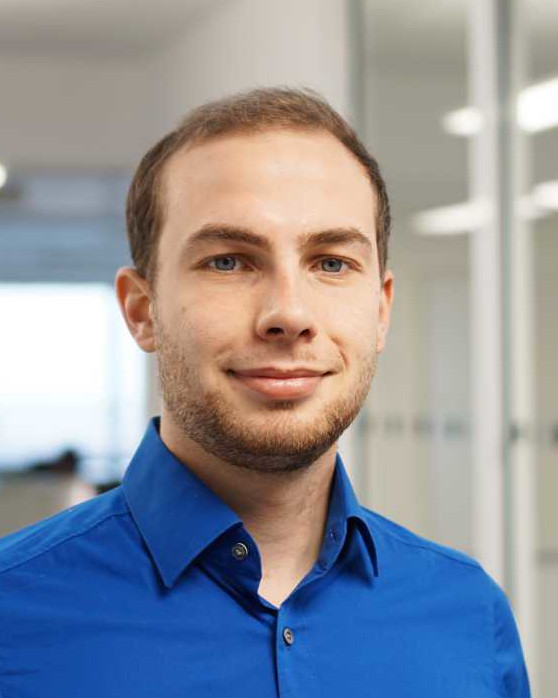}}]{Leonardo Mörlein}
Leonardo Mörlein (Graduate Student Member, IEEE) was born in 1994 in Würzburg, Germany. He received the B.Sc. and M.Sc. degrees in electrical engineering from Leibniz University Hannover, Hannover, Germany, in 2017 and 2020, respectively. He is currently a Research Assistant with the Institute of Microwave and Wireless Systems, Leibniz University Hannover. His current research focuses on the use of multi-port multi-mode antennas in beamforming scenarios. Further research interests include antenna integration, the use of modal decompositions and channel modeling.
\end{IEEEbiography}

\begin{IEEEbiography}[{\includegraphics[width=1in,height=1.25in,clip,keepaspectratio]{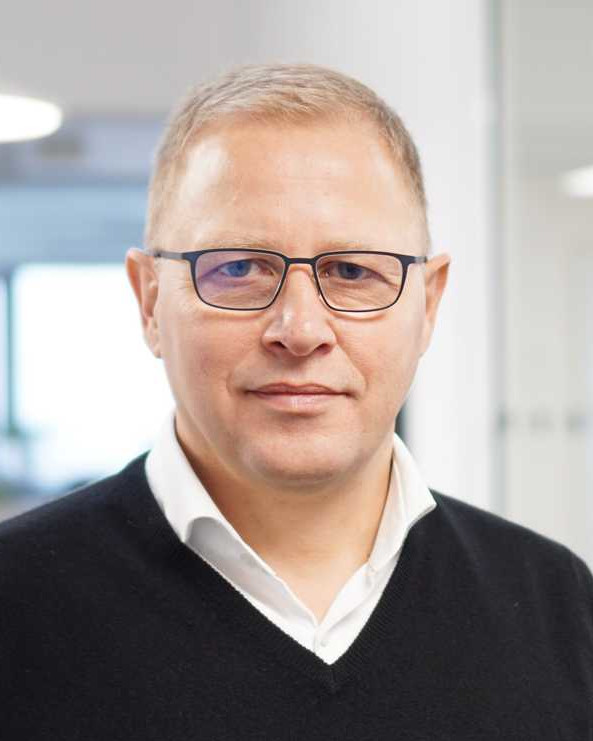}}]{Dirk Manteuffel} (Member, IEEE) was born in Issum, Germany, in 1970. He received the Dipl.-Ing. and Dr.-Ing. degrees in electrical engineering from the University of Duisburg–Essen, Duisburg, Germany, in 1998 and 2002, respectively.

From 1998 to 2009, he was with IMST, Kamp-Lintfort, Germany. As a Project Manager, he was responsible for industrial antenna development and advanced projects in the field of antennas and electromagnetic (EM) modeling. From 2009 to 2016, he was a Full Professor of wireless communications at Christian-Albrechts-University, Kiel, Germany. Since June 2016, he has been a Full Professor and the Executive Director of the Institute of Microwave and Wireless Systems, Leibniz University Hannover, Hannover, Germany. His research interests include electromagnetics, antenna integration and EM modeling for mobile communications and biomedical applications.

Dr. Manteuffel was a director of the European Association on Antennas and Propagation from 2012 to 2015. He served on the Administrative Committee (AdCom) of IEEE Antennas and Propagation Society from 2013 to 2015 and as an Associate Editor of the IEEE Transactions on Antennas and Propagation from 2014 to 2022. Since 2009 he has been an appointed member of the committee "Antennas" of the German VDI-ITG.
\end{IEEEbiography}





\end{document}